\documentclass[acmtog]{acmart}
\setcitestyle{nosort,square} %

\usepackage{xcolor, colortbl}
\usepackage{tikz}
\usetikzlibrary{shapes.geometric, arrows}

\usepackage{cuted}
\usepackage{tabularx}

\acmJournal{TOG}

\setcopyright{rightsretained}
\acmJournal{TOG}
\acmYear{2022}\acmVolume{41}\acmNumber{4}\acmArticle{119}\acmMonth{7} 
\acmDOI{10.1145/3528223.3530169}

\usepackage{hyperref}
\usepackage{url}
\usepackage{wrapfig} 
\usepackage{multirow}
\usepackage{overpic}
\usepackage{subcaption}

\definecolor{purple}{rgb}{0.5,0.0,0.97}
\definecolor{cyan}{rgb}{0.0,0.6,0.8}
\definecolor{green}{rgb}{0.3,0.56,0.0}

\newcommand{\revise}[1]{#1}
\newcommand{\myrefeq}[1]{Eq.~\ref{#1}}
\newcommand{\myreffig}[1]{Fig.~\ref{#1}}
\newcommand{\myreftab}[1]{Table~\ref{#1}}
\newcommand{\myrefsec}[1]{Sec.~\ref{#1}}

\newcommand{\resizeEq}[3]{
\begin{equation}%
\resizebox{#3}{!}{%
$\begin{array}{@{}cc}
\begin{split}
#1
\end{split}
\end{array}$}  \label{#2}
\end{equation}
}

\begin{document}
\title{Physics Informed Neural Fields for Smoke Reconstruction with Sparse Data}

\author{Mengyu Chu}
\email{mchu@mpi-inf.mpg.de}
\affiliation{
\institution{Max Planck Institute for Informatics, SIC}%
\country{Germany}
}
\author{Lingjie Liu}
\email{lliu@mpi-inf.mpg.de}
\affiliation{
\institution{Max Planck Institute for Informatics, SIC}%
\country{Germany}
}
\author{Quan Zheng}
\email{zhengquan@iscas.ac.cn}
\affiliation{
\institution{Institute of Software Chinese Academy of Sciences}%
\country{China}
}
\author{Erik Franz}
\email{franzer@in.tum.de}
\affiliation{
\institution{Technical University of Munich}
\country{Germany}
}
\author{Hans-Peter Seidel}
\email{hpseidel@mpi-sb.mpg.de}
\affiliation{
\institution{ Max Planck Institute for Informatics, SIC}
\country{Germany}
}
\author{Christian Theobalt}
\email{theobalt@mpi-inf.mpg.de}
\affiliation{
\institution{ Max Planck Institute for Informatics, SIC}
\country{Germany}
}
\author{Rhaleb Zayer}
\email{rzayer@mpi-inf.mpg.de}
\affiliation{
\institution{ Max Planck Institute for Informatics, SIC}
\country{Germany}
}

\renewcommand\shortauthors{Chu, M.; Liu, L.; Zheng, Q.; Franz, E.; Seidel, HP.; Theobalt C.; Zayer R.}

\begin{abstract}
High-fidelity reconstruction of dynamic fluids from sparse multiview RGB videos remains a formidable challenge, due to the complexity of the underlying physics as well as the severe occlusion and complex lighting in the captured data. Existing solutions either assume knowledge of obstacles and lighting, or only focus on simple fluid scenes without obstacles or complex lighting, and thus are unsuitable for real-world scenes with unknown lighting conditions or arbitrary obstacles. We present the first method to reconstruct dynamic fluid phenomena by leveraging the governing physics (ie, Navier -Stokes equations) in an end-to-end optimization from a mere set of sparse video frames without taking lighting conditions, geometry information, or boundary conditions as input. Our method provides a continuous spatio-temporal scene representation using neural networks as the ansatz of density and velocity solution functions for fluids as well as the radiance field for static objects. With a hybrid architecture that separates static and dynamic contents apart, fluid interactions with static obstacles are reconstructed for the first time without additional geometry input or human labeling. By augmenting time-varying neural radiance fields with physics-informed deep learning, our method benefits from the supervision of images and physical priors.
To achieve robust optimization from sparse input views, we introduced a layer-by-layer growing strategy to progressively increase the network capacity of the resulting neural representation.
Using our progressively growing models with a newly proposed regularization term, we manage to disentangle the density-color ambiguity in radiance fields without overfitting. 
A pretrained density-to-velocity fluid model is leveraged in addition as the data prior to avoid suboptimal velocity solutions which underestimate vorticity but trivially fulfill physical equations. Our method exhibits high-quality results with relaxed constraints and strong flexibility on a representative set of synthetic and real flow captures.
Code and sample tests are at 
\url{https://people.mpi-inf.mpg.de/~mchu/projects/PI-NeRF/}.
\end{abstract}



\begin{teaserfigure}
	\centering
	{\setlength{\fboxsep}{0.02pt}\colorbox{black}{		\begin{minipage}{0.6\textwidth}
	\scalebox{-1}[1]{\includegraphics[ height=96pt]{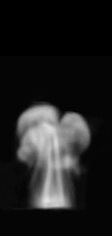}}
	\includegraphics[height=90pt]{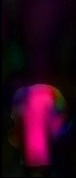}
	\includegraphics[ height=90pt]{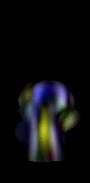}
	\scalebox{-1}[1]{\includegraphics[ height=96pt]{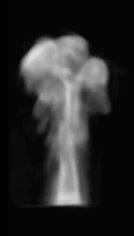}}
	\includegraphics[ height=90pt]{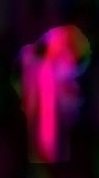}
	\includegraphics[ height=90pt]{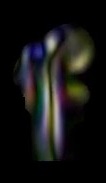}
	\newline
	\begin{overpic}[height=90pt]{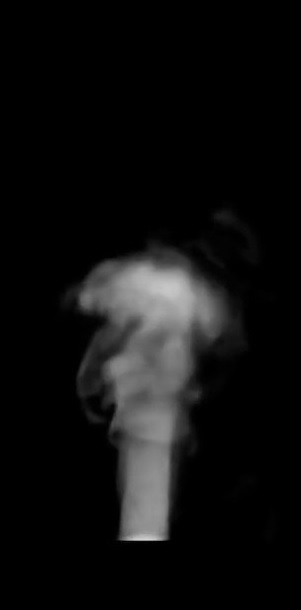}
	\put(3,182){{\color{yellow!80}  - Synthetic plume -}}
	\put(5,102){{\color{white}  \footnotesize  \textit{Novel view}}}
	\put(8,92){{\color{white}  \footnotesize  \textit{rendering}}}
	\end{overpic}
	\begin{overpic}[height=92pt]{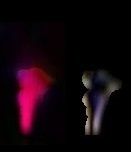}
	\put(8,102){{\color{white}  \footnotesize  \textit{Velocity and vorticity}}}
	\put(16,92){{\color{white}  \footnotesize  \textit{visualizations}}}
	\put(-48,2){{\color{yellow!80}  - Real capture -}}
	\end{overpic}
	\begin{overpic}[height=92pt]{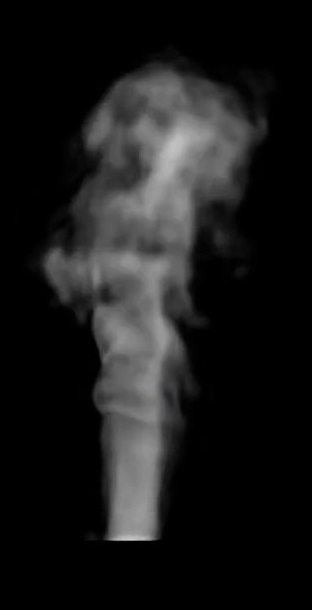}
	\put(20,102){{\color{white}  \footnotesize  \textit{Rendering,
	velocity and vorticity at a later time step}}}
	\end{overpic}
	\includegraphics[ height=92pt]{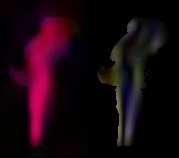}
	\end{minipage}
	\hfill
	\begin{minipage}{0.3\textwidth}
	\begin{overpic}[height=189pt]{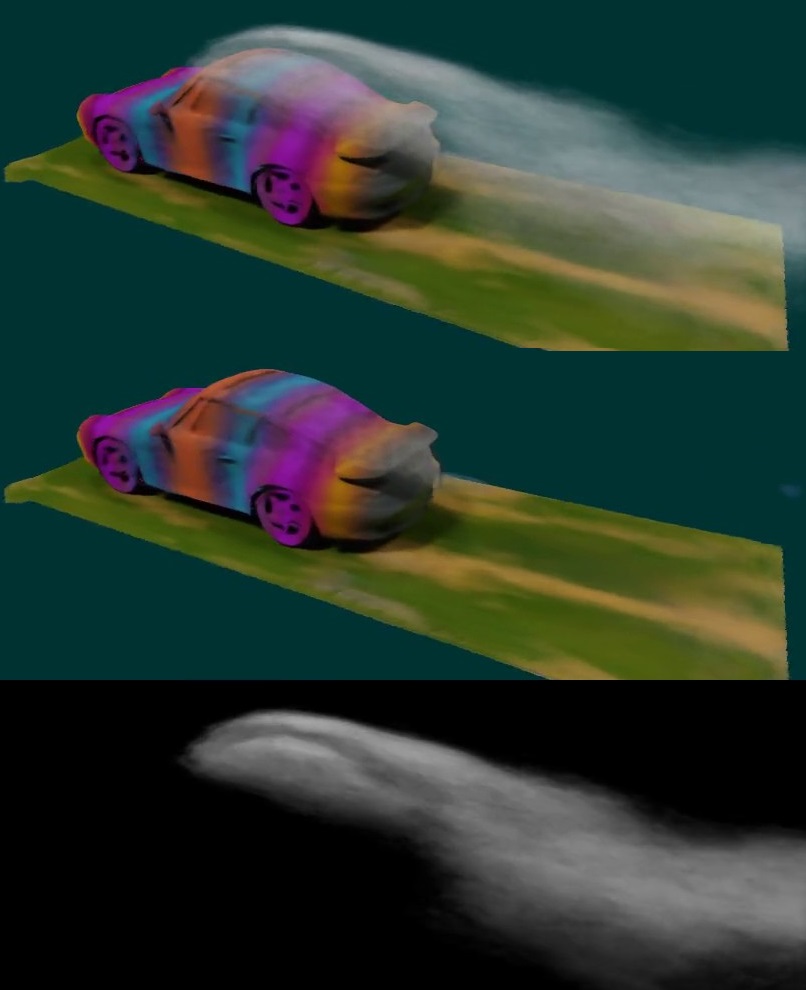}
	\put(2,2){{\color{yellow!80}  - Hybrid Scene -}}
	\put(38,28){{\color{white}  \footnotesize \textit{Rendering of the dynamic part}}}
	\put(42,60){{\color{white}  \footnotesize \textit{Rendering of the static part}}}
	\put(56,96){{\color{white}  \footnotesize \textit{Rendering of both}}}
	\end{overpic}
	\end{minipage}}}
	\vspace{-12pt}
	\caption{\footnotesize
	Renderings and visualizations of fluid reconstructions. Using sparse image sequences, we learn continuous radiance and velocity fields to represent fluid scenes.
	We can handle synthetic and real scenes (on the left) as well as hybrid scenes (on the right). Our method supports arbitrary obstacles and unknown lighting conditions flexibly.}
	\label{fig:teaser} %
\end{teaserfigure}

\maketitle
\section{Introduction}

Obeying general laws of physics, fluid phenomena are ubiquitous and various.
The understanding of fluids benefit a wide range of activities including weather forecasting~\cite{bauer2015quiet}, vehicle manufacturing~\cite{bushnell1991drag}, medicine metabolism~\cite{shi2010quantitative}, and visual effects~\cite{kim2008wavelet,you2018tempoGAN}.
In studying the fluid behaviors, one of the core tasks is to estimate the invisible velocity.
In general, velocity can be obtained either by solving physical equations with numerical solvers~\cite{bridson2015fluid} or by measuring experimentally, e.g. particle image velocimetry (PIV)~\cite{elsinga2006tomographic}, both with different pros and cons.

With the progress in hardware and algorithms, numerical solvers achieve high accuracy in ``forward'' fluid simulation tasks, e.g. analyzing canonical flows, where experienced engineers are able to provide a full set of initial and boundary conditions, etc.
However, due to partially available initial conditions, common users cannot easily apply them in an ``inverse'' manner to handle particular fluid phenomena in real life, e.g. steam rising up from a teapot.
Experimental techniques, including PIV, allow users to estimate real-life fluid behavior. In these works~\cite{Grant97PIV, xiong2017rainbow}, the scope is limited to simple scenes with fluid as the main objective. 
By estimating the volumetric distribution of a passive scalar, e.g. tracer particles, colorful dye, or smoke, from image modalities, 
the underlying motion is then inferred from the transport and physical equations.
As PIV methods require specialized lab settings, there has been growing interest in fluid reconstructions from RGB images~\cite{gregson2014capture, franz2021global}.
Despite improvements in reconstruction quality and reduction of hardware, existing approaches are still vulnerable to scenes with unknown illumination or occluding obstacles.

We aim to overcome these difficulties, handle fluid scenes with unknown lighting and arbitrary obstacles, and advance the goal of capturing fluids in the wild. 
In addition to reducing the constraints for fluid capture, being able to handle fluid scenes with obstacles is an important step towards analyzing fluid-obstacle interactions.

Based on recent progress in view synthesis using neural radiance fields (NeRF)~\cite{lombardi2019neural,mildenhall2020nerf}, %
we first augment the spatial scene representation in the time dimension and learn a time-varying neural radiance field.
Taking RGB videos of a scene with dynamic fluid as input, the time-varying NeRF learns to encode a spatiotemporal radiance field with a Multi-Layered Perceptron (MLPs).
Originally proposed for static scenes, NeRF uses differentiable rendering to gather information from multiview images with different camera poses.
Correspondingly, in our case, it is important to apply differentiable physics to unify information from video frames at different time steps as a dynamic NeRF.
Thus, we propose to use physics-informed deep learning technologies~\cite{raissi2020hidden} and use another MLP to represent the continuous spatio-temporal velocity field.
With jointly applied differentiable rendering
and physics models,
we manage to learn the continuous spatiotemporal radiance and velocity fields in a dynamic scene and supervise them using image sequences and physical laws end to end.
Similar to Physics-Informed Neural Networks (PINN), partial differential equations (PDE) are calculated using the exact, mesh-free derivatives of the velocity model via auto-differentiation without discretization.

\begin{figure}
    \centering
    \includegraphics[width=0.8\linewidth]{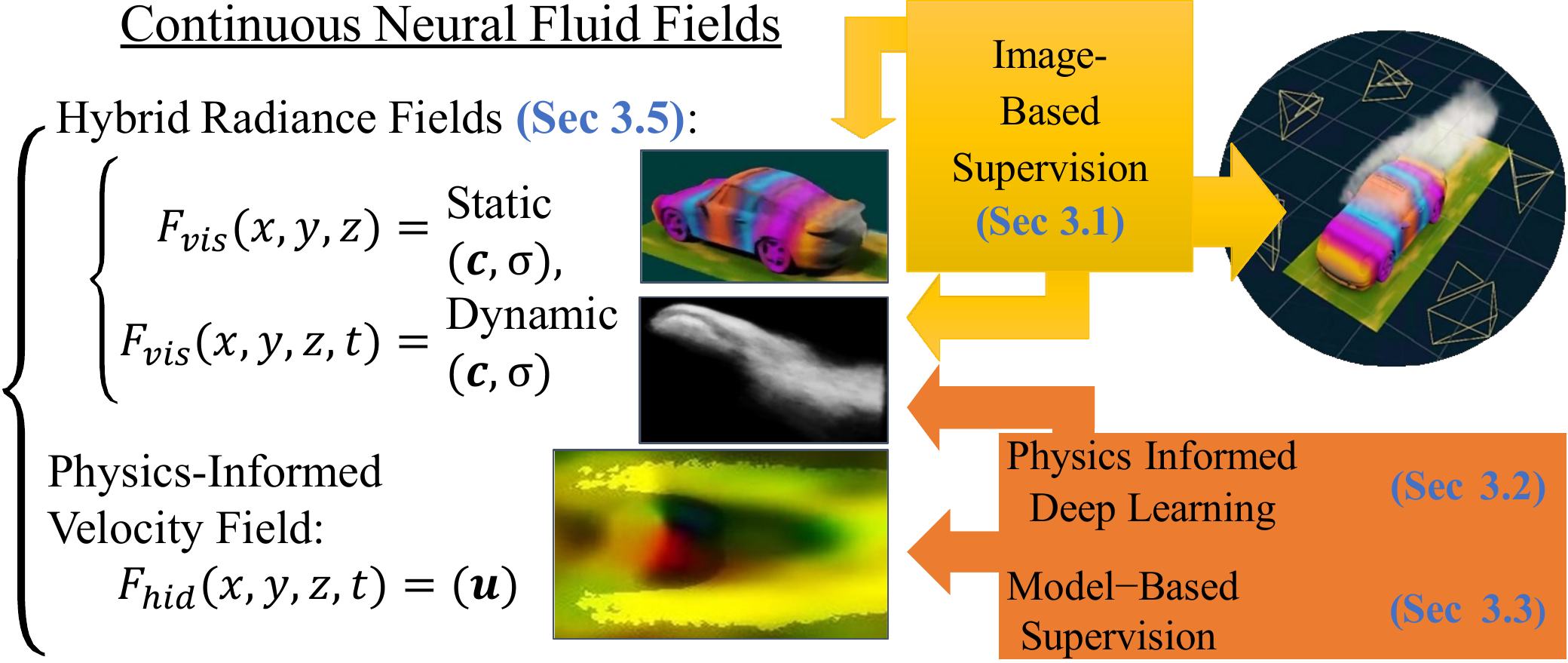}
    \vspace{-12pt}
    \caption{\footnotesize 
    Modules and supervision in our method.
    With the comprehensive supervision of images, physical-priors, and a published fluid network, we present neural fluid fields with radiance fields for static and dynamic components, and velocity fields for fluids.}
    \label{fig:outline}\vspace{-12pt}
\end{figure}

Through a seamless intertwining of NeRF and PINN,
our method learns physics-informed neural fluid fields from image sequences end to end. 
It is unrestrained by lighting conditions, geometry information, or the initial and boundary conditions, and is therefore suitable for hybrid scenes with obstacles. 
Our method also face the same issues of PINN methods: the training task is challenging due to the complex non-linear optimization landscape shaped by PDE constraints. 
Specifically, in fluid reconstruction, this issue tends to manifest itself as underestimated vorticity and density overfitting to given views. 
While overfitting problems are usually handled with regularization terms, more data, or model-based supervision,
we propose to use a regularization term on the density and a published pre-trained velocity model~\cite{chu2021learning} to reduce the vorticity underestimation.
While both Neural Volumes~\cite{lombardi2019neural} and the static NeRF have to use a large number of images with different camera poses to properly disentangle the radiance color and opacity, we show that our density regularization term helps to effectively disentangle this color-density ambiguity of the radiance field, which eventually allows us to work with sparse camera views. 
The general pipeline of our algorithm is illustrated in \myreffig{fig:outline}.

To summarize, our work makes the following contributions: %
\begin{itemize}\vspace{-3pt}
    \item We propose the first method for reconstructing dynamic fluids with high quality from sparse multi-view RGB videos, without access to any lighting and geometry information. 
    \item We introduce a hybrid representation for dynamic fluid scenes interacting with static obstacles. It allows automatic separation of dynamic and static components without any human labelling, which is previously not possible. 
    \item We propose comprehensive supervision using images, physical priors (i.e., Navier-Stokes equations), as well as a pre-trained fluid model as data prior to prevent sub-optimal solutions with underestimated vorticity.
    \item We propose a progressively growing model with a new regularization term to penalize ``ghost density'', an artifact caused by color-density ambiguity.Together, overfitting to sparse given views is avoided.
\end{itemize}\vspace{-2pt}
With these, our method relaxes the requirements of fluid reconstruction, enables the capture of fluid behavior with obstacles, and achieves state-of-the-art results in both synthetic and real scenes. 
\section{Related Work}
We summarize related work in fluid reconstruction, physics-informed deep learning, and neural scene representations.

\subsection{Fluid Reconstruction Methods} 
Several approaches have been proposed to reconstruct fluids from observations of visible light measurements.
There are established methods that use active sensing with specialized hardware and lighting setups~\cite{Atcheson2008OEF,Gu13StructuredLight,Ji2013LightPath} and particle imaging velocimetry (PIV)~\cite{Grant97PIV,elsinga2006tomographic,xiong2017rainbow} injecting passive markers into fluid flows.
PIV methods usually require specialized setups for particles, lighting, and capture.
Many flow phenomena, e.g. smoke and fire, have visible elements which can easily be recorded.
In the following, we focus on fluid reconstruction using RGB images to alleviate the need for specialized setups.
Early work~\cite{gregson2014capture} uses linear image formation to  
extract passive quantities from simple RGB images and reconstruct fluids with physical priors.
Extending this direction, ScalarFlow~\cite{eckert2019scalarflow} reconstructs real-world smoke plumes from sparse views and \citet{zang2020TomoFluid} use interpolated views as further constraints.
Instead of using linear image formation, the Global Transport method~\cite{franz2021global} uses differentiable rendering to allow end-to-end optimization. %
\citet{qiu2021RapidEndtoend} train convolutional networks end-to-end to estimate flow from sequences of orthogonal views.
These methods either require known lighting conditions or ignore lighting altogether. 
Most related work reconstruct velocity on discrete grids, while we employ MLPs as an ansatz for fluid and present continuous velocity fields.

\subsection{Physics Informed Deep Learning}
Physics-informed deep learning respects physical equations governing a certain problem, e.g. Partial Differential Equations (PDE), by coupling them into the learning process.
These methods~\cite{PAKRAVAN2021110414, RAISSI2019} have emerged as essential tools for various challenging forward and inverse problems.
By coupling differentiable numerical solvers into training, physically plausible solutions can be obtained for dynamic problems, e.g. cloths~\cite{geng2020coercing} and fluids~\cite{um2020solver, gibou2019sharp}.
Besides using discretized numerical solvers, PINN~\cite{RAISSI2019} proposes employing MLPs as continuous network surrogates for dynamics.
Early MLP-based models~\cite{Lagaris98,He00,MAIDUY2003} use a small number of hidden layers and hyperbolic tangent or sigmoid nonlinearities. 
The network capacity is restricted. 
Current approaches~\cite{Sirignano2018,raissi2020hidden,BERG2018} advanced on equation sophistication and dimensionality by capitalizing on new optimization frameworks and autodifferentiation for training MLP-based networks. 
Recent studies~\cite{sitzmann2020implicit, tancik2020fourfeat} have demonstrated that the commonly used MLPs struggle with high-frequency information and propose new periodic activation functions to represent complex natural signals and their derivatives.

\subsection{Neural Representations} 

In scene representations, neural representations have recently been widespread for their expressiveness and compactness.

\paragraph{Neural intermediate representations} To synthesize novel views of a 3D scene, multi-plane images~\cite{zhou2018stereo,mildenhall2019local,srinivasan2019pushing}, multi-sphere images~\cite{broxton2020immersive} and proxy geometries~\cite{zhang2021neural,philip2019multi} are proposed to build intermediate neural representations from multiple views. Novel views can be rendered with learned representations in interpolation or extrapolation manners. \citet{eslami2018neural} propose to learn a neural embedding from observed images of a scene and infer unseen novel views. \citet{granskog2020compositional} further partition and compress the neural embedding into concise components which enables compositional rendering.

\paragraph{Neural explicit representations} By leveraging inverse rendering, classic 3D representations have been adapted to neural versions. \citet{yifan2019differentiable} introduce differentiable surface splatting to enable point-cloud-based geometry processing from images. DeepVoxels~\cite{sitzmann2019deepvoxels} learns a 3D feature representation and stores features in a voxel grid for novel view synthesis. Neural Volumes~\cite{lombardi2019neural} encodes objects in a voxel grid and learns an inverse mapping to decode voxelized radiance. These grid-based discretization suffers from resolution limitations.

\paragraph{Neural implicit representations} To avoid the resolution limitation of discretization, implicit 3D representations, including signed distance functions (SDF)~\cite{park2019deepsdf, chabra2020deep}, unsigned distance functions~\cite{chibane2020neural}, and occupancy fields~\cite{chen2019learning, peng2020convolutional}, are parameterized with neural networks to represent 3D shapes. Furthermore, these continuous representations are generalized to coordinate-based networks. For 3D shapes with textured surfaces, scene representation networks~\cite{sitzmann2019scene} learn SDFs and texture color for each coordinate. \citet{niemeyer2020differentiable} derive implicit gradients to enable optimization and learn the surface radiance. Instead of SDF, PIfu~\cite{saito2019pifu} proposes to predict surface occupancy and color for coordinates. 
Different from scene representations based on opaque surfaces, NeRF~\cite{mildenhall2020nerf} represents scenes as implicit volumes and trains the coordinate-based networks to approximate continuous volumetric radiance fields. After training, novel views are rendered by ray marching. Many extensions have been proposed for fast rendering~\cite{liu2020neural,Reiser2021ICCV,yu2021plenoctrees,garbin2021fastnerf,hedman2021snerg}, geometry reconstruction~\cite{yariv2020multiview,Oechsle2021ICCV,yariv2021volume,wang2021neus}, generative image synthesis~\cite{schwarz2020graf,Niemeyer2020GIRAFFE,gu2021stylenerf}, reflectance for opaque surfaces~\cite{srinivasan2020nerv}, and participating media reconstruction~\cite{zheng2021neural}.%

\subsection{Neural Representations for Dynamic Scenes}
As the above representations apply mainly to static scenes, there has been a steady effort targeting representations for dynamic scenes. 
Neural Volumes~\cite{lombardi2019neural} and its follow-up works~\cite{wang2020learning,lombardi2021mixture,raj2021pva} employ an encoder-decoder to transform input images to a volume representation, followed by differentiable ray-marching. 
Some works~\cite{tretschk2020nonrigid,park2020nerfies,Li2020NeuralSF,park2021hypernerf} design a dedicated deformation network to model the dense 3D motions of points between adjacent frames. \citet{xian2020spacetime} model dynamic scenes as 4D space-time irradiance fields. 
Several works focus on human body rendering. ~\citet{gafni2020dynamic} condition NeRF on face pose to represent a 4D facial avatar. 
Neural Body~\cite{peng2020neural} uses a SMPL model as a 3D proxy and attaches learnable features on each SMPL mesh vertices as anchors to connect spaces across difficult frames.
\citet{su2021anerf} present an articulated NeRF based on a human skeleton for refining human pose estimation. 
Some works~\cite{liu2021neural,peng2021animatable,chen2021animatable} propose a geometry-guided deformable NeRF method for warping the space in different poses to a shared canonical space with a SMPL model as a 3D proxy. Neural Actor~\cite{liu2021neural} learns  pose-dependent deformation and appearance to model dynamic effects. Neural Human Performer~\cite{kwon2021neural} proposes a generalizable NeRF based on a SMPL model with a temporal transformer and a multi-view transformer for new pose and appearance synthesis. 
In contrast, we focus on modeling the dynamic fluid with a new neural representation incorporated with physics-based constraints, which has not received much attention yet.

\section{Neural Scene Representation For Fluids}\label{sec:method}

Based on the universal approximation theorem, our method uses neural networks to represent a fluid scene in space and time.
Mathematically, we use two MLP-based networks, 
\resizeEq{
F_{vis} :(x,y,z,t)  \rightarrow  (\mathbf{c},\sigma) \ \text{and} \ 
F_{hid} :(x,y,z,t)  \rightarrow  (\mathbf{u})
}{eq:mlps}{!}
to approximate the continuous functions of the radiance field and velocity field, respectively.
The input, $(x,y,z,t)$, is the coordinate of a 4D location in space and time,
the radiance output of $F_{vis}$ contains the emitted color $\mathbf{c}=(c_r,c_g,c_b)$ and the optical density $\sigma$.
The velocity output of $F_{hid}$ describes the 3D motion vector $\mathbf{u}=(u_x,u_y,u_z)$ at that point and time.
\revise{``\textit{hid}'' stands for the word ``hidden'', since the velocity field is not directly observed.}
As a representation for fluid scenes, the radiance field provides us useful information about the lighting condition, object geometry, and the density distribution of the passive scalar in the scene. Meanwhile the velocity field describes the invisible dynamics which is important for further analysis of pressure or body forces for fluid mechanics.
While it is possible to supervise $F_{vis}$ directly using video sequences based on differentiable volumetric rendering, $F_{hid}$ has to be trained indirectly from the density distribution of the passive scalar via physics equations.
Note that physics equations are associated with mass density $d$, which can be considered as being proportional to the optical density $\sigma$ according to the Beer–Lambert law. 

In the following, we describe the image-based supervision of $F_{vis}$ in \myrefsec{sec:f_visible} and the physical equation-based supervision of $F_{hid}$ in \myrefsec{sec:f_hidden}. 
Facing the highly non-linear and challenging optimization landscape formed by PDEs, we propose a model-based supervision in \myrefsec{sec:f_hidden_model} to avoid sub-optimal  solutions of $F_{hid}$ with underestimated vorticity.
In order to obtain a valid density distribution which will determine the best possible accuracy of $F_{hid}$, 
we propose to tackle the color-density ambiguity (\myrefsec{sec:col-den}) and disentangle the fluid and obstacle density (\myrefsec{sec:hybrid}), respectively.

\subsection{Image-Based Radiance Estimation}\label{sec:f_visible}

Briefly summarized, the static NeRF~\cite{mildenhall2020nerf} learns a static neural radiance field  $F_{NeRF} : (x,y,z,\theta,\phi) \rightarrow (\mathbf{c},\sigma)$ with a set of posed images though volumetric rendering. Here, $(x,y,z)$ is the coordinate of a position in 3D space and ($\theta$, $\phi$) defines a ray direction as a 3D Cartesian unit vector $\mathbf{d}$.
Considering the pixel $j$ of image $i$,
point samples are queried along the camera ray $r_{ij}(h) = \mathbf{o}_i + h \mathbf{d}(\theta_{ij}, \phi_{ij})$, where $h$ denotes the parametric distance. The color of the pixel is approximated with the numerical quadrature rule:
\resizeEq{
\mathbf{C}(r_{ij}) = &  \sum_{k=1}^{K} T(h_k) \alpha(h_k) \mathbf{c}(h_k), & \text{~where~} &
T(h_k) =  exp\Big(-\sum_{\hat{k}=1}^{k-1} \sigma(h_{\hat{k}}) \delta_{\hat{k}} \Big)\text{~,~}  \\
\alpha(h_k) = & 1 - exp(-\sigma(h_k) \delta_{k}), & \text{~and~} &  \delta_{k} =  h_{k+1} - h_{k} \text{~.~}  \\
}{eq:nerf}{0.9\linewidth}
To sample rays efficiently with adaptive ray step $\delta_{k}$, NeRF simultaneously optimizes two MLPs, a ``coarse'' one as a probability density function for importance sampling and a ``fine'' one as the target radiance field being sampled.
Their parameters are then optimized with the following objective functions:
\resizeEq{
\mathcal{L}_{img} = 
\sum_{ij} \| \mathbf{C}_{img}(r_{ij}) - \mathbf{C}_c(r_{ij})\|_2^2 + \| \mathbf{C}_{img}(r_{ij}) - \mathbf{C}_f(r_{ij})\|_2^2,
}{eq:nerf_l2}{0.9\linewidth}
where $\mathbf{C}_{img}(r_{ij})$ is the reference color of pixel $j$ in image $i$, and $\mathbf{C}_c$ and $\mathbf{C}_f$ are the color rendered by the coarse and fine models, respectively.
Based on the ray casting based rendering algorithm, NeRF establishes the connection between 2D images and 3D geometry, supporting both solid objects as well as volumetric media.

\begin{wrapfigure}{R}{0.35\linewidth}
  \begin{center}%
    \includegraphics[width=\linewidth]{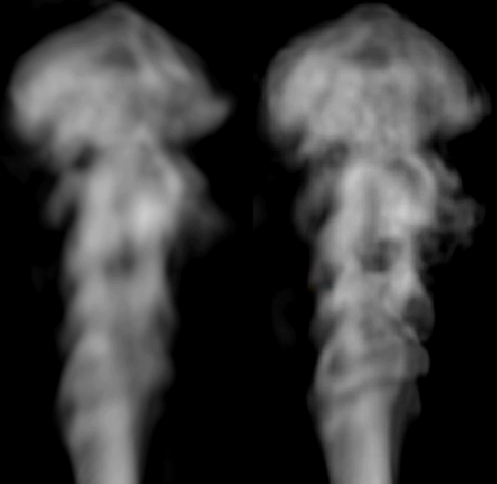}\\
    {\footnotesize 
    a) w.o. \myrefeq{eq:nerf_vgg} \hspace{0.1\linewidth}
    b) with \myrefeq{eq:nerf_vgg}
    }
  \end{center}
  \caption{\footnotesize The VGG loss helps improve the perceptual quality of the reconstructed results and capture more high frequency details.} \label{fig:vgg}
\end{wrapfigure}
In order to deal with dynamic scenes of fluid and explore the temporal evolution, we perform the following adaptations.
First, we extend the static NeRF model with time $t$ as input.
Second, we assume that the dynamic scene to be reconstructed only consists of Lambertian surfaces and media with isotropic scattering. Therefore, we remove the input $\theta$ and $\phi$ for simplicity.
Finally, we use the MLP with periodic activation functions proposed in SIREN~\cite{sitzmann2020implicit} instead of ReLU-based MLPs with positional encoding strategies in order to model the continuous derivatives better.
Same as the original NeRF, the quadrature rule is used for volumetric rendering and two models are trained as hierarchical volume sampling.
Besides the $L_2$ loss in \myrefeq{eq:nerf_l2},
we include a VGG-based perceptual loss term as proposed in previous work on image and video synthesis~\cite{ledig2017photo, chu2020learning}.
Since it is time-consuming to render all pixels of a given image using the $F_{vis}$ model, 
we randomly crop square patches from images with varying strides.
In this way, only small image patches in resolution of $40\times40$ are rendered during the training and we get VGG features
at different scales. The VGG feature distances are measured using cosine similarity by:
\resizeEq{
\mathcal{L}_{VGG} =  \sum_{\phi \in \text{\footnotesize VGG layers}} 1.0 - \frac {\Phi(\text{I}_{img}) * \Phi(\text{I}_{vis}) }{\left \| \Phi(\text{I}_{img})\right \| * \left \|\Phi(\text{I}_{vis})\right \|}
\text{~, with~} \\
\text{I}_{vis} = \{\mathbf{C}_{c \text{~or~} f}(r_{ij}) | {j \in [j_y:j_y+40s:s,j_x:j_x+40s:s ]} \}\\
\text{I}_{img} =  \{\mathbf{C}_{img}(r_{ij}) | {j \in [j_y:j_y+40s:s,j_x:j_x+40s:s ]} \}.
}{eq:nerf_vgg}{!}
As shown in \myreffig{fig:vgg}, supervising the radiance fields with $l_2$ and VGG feature losses together helps to capture detailed structures better. 

With these changes, we get a time-varying NeRF model based on SIREN layers: {$F_{vis} : (x,y,z,t) \rightarrow (\mathbf{c},\sigma)$ }.
It models the continuous functions of density and color and their derivatives in the space and time.
Continuous derivative is necessary for learning the temporal evolution, which will be explained in the following section.

\subsection{Physics-Informed Velocity Estimation}\label{sec:f_hidden}

Recent approaches~\cite{RAISSI2019, raissi2020hidden} demonstrate that deep-learning models can be trained as data-driven solutions of physical problems via optimizing the governing PDEs, which are known as PINN.
A distinctive feature shared by these studies is to use the continuous and mesh-free derivatives computed with auto-differentiation.
For fluid dynamics,
\citet{raissi2020hidden} train neural networks, {$F_{fluid} : (x,y,z,t) \rightarrow (d,\mathbf{u},p)$ },
as the ansatz of the underlying solution functions.
They propose to supervise the network using a ground-truth spatio-temporal dataset describing the  distribution of the passive scalar with discrete data pairs $\{<(x,y,z,t), d>\}_n$.
The aforementioned auto-differentiation is used to optimize the transport equation: $\frac{\partial d}{\partial t} +\mathbf{u} \cdot \nabla d = 0$,
as well as the Navier-Stokes equations:
\resizeEq{
	\frac{\partial \mathbf{u}}{\partial t} +\mathbf{u} \cdot \nabla \mathbf{u} & = 
	- \frac{1}{\rho} \nabla p
	+ \nu \nabla \cdot \nabla \mathbf{u} 
	+ \mathbf{f} \ , \\
	\nabla \cdot \mathbf{u} & = 0 \ .
}{eq:nse}{!}
Without requiring boundary conditions as input, 
this method can handle fluid taking place in arbitrarily complex spacial domains.
High accuracy of velocity $\mathbf{u}$ and pressure $p$ can be achieved with a shallow MLP when physical parameters, e.g. kinematic viscosity $\nu$, are given and the density $d$ is densely sampled from exact solutions.

Based on the PINN technology, we propose to learn a velocity network, {$F_{hid} : (x,y,z,t) \rightarrow \mathbf{u}$ }, while training the radiance fields network {$F_{vis} : (x,y,z,t) \rightarrow (\mathbf{c},\sigma)$ } mentioned in \myrefsec{sec:f_visible}.
Same as in $F_{vis}$, $F_{hid}$ consists of SIREN layers~\cite{sitzmann2020implicit} with a principled initialization scheme, which allows for training deeper networks. Compared to the learning from ground-truth data~\cite{raissi2020hidden}, we are facing a tougher learning task with more ambiguity since the density distribution $<(x,y,z,t), d>$ is not accurately provided by the training data, but is represented by $\sigma$ of $F_{vis}$ which is simultaneously optimized from images through volumetric rendering.
We optimize our velocity model $F_{hid}$ by minimizing:
\resizeEq{
\mathcal{L}_{\frac{D\sigma}{Dt}} & = \Big(\frac{\partial \sigma}{\partial t} +\mathbf{u} \cdot \nabla \sigma \Big)^2 \text{~,~and~}\\
\mathcal{L}_{NSE} & = 
	\Big\| \frac{\partial \mathbf{u}}{\partial t}+\mathbf{u} \cdot \nabla \mathbf{u}\Big\|_2^2  + w_{div}  \Big\|
	\nabla \cdot \mathbf{u}\Big\|_2^2 \ .
}{eq:nseloss}{!}
Instead of optimizing additional pressure fields and extra forces with largely increased degrees of freedom, we have dropped out the right-hand side in \myrefeq{eq:nse} as a simplification with valid assumptions, which is equivalent to searching for a possible solution with minimal influence from extra forces, pressure difference, and viscosity.

\myrefeq{eq:nseloss} is used to train $F_{hid}$ as a strong physics-based prior.
It cannot be applied to $F_{vis}$, otherwise $\sigma$ will be trivially reduced simply to minimize $\mathcal{L}_{\frac{D\sigma}{Dt}}$.
To improve the temporal consistency of the radiance field with physics-informed learning, we propose to optimize the radiance field across time with warping.
Considering the volumetric rendering of \myrefeq{eq:nerf}, instead of tracing rays at the original positions $r^{*}_{ij}(h) = o_i + h\mathbf{d}$ for pixel $j$ in image $i$ with a time-step $t$, we query radiance with point samples at warped positions, i.e.:
\resizeEq{
\Big(\mathbf{c}(h),\sigma(h)\Big) =  F_{vis}\Big(r_{ijt}(h), t+\delta t\Big) \ , \\
r_{ijt}(h) = r^{*}_{ij}(h) + \mathbf{u}(h,t) \delta t\ , \ 
\mathbf{u}(h,t)= F_{hid}\Big(r^{*}_{ij}(h), t\Big)
}{eq:warp}{!}
with $\delta t \sim \mathcal{N}(0,0.5)$,
which stochastically associates the current frame with its previous and next frames at $t-1$ and $t+1$.
This is in line with the ray-bending used in Non-Rigid NeRF~\cite{tretschk2020nonrigid} and the flow warping in NeuralSF~\cite{Li2020NeuralSF}, but in a stochastic form which is more appropriate for turbulent fluid motion than frame-to-frame warping with discrete time steps. 
\revise{ %
We use the Euler method to calculate warped positions for efficiency.
Higher-order methods, e.g. Runge-Kutta methods, can provide better accuracy. This will better preserve temporal consistency and spacial detail, but requires longer computation time due to multiple velocity queries.
}
Note that we only warp the density field but use the original color, since the color is not a transported scalar but an attribute resulting from lighting and density distribution. The corresponding rendering objective function with warping is denoted as $\mathcal{L}_{\widetilde{img}}$ in the rest of the paper.

\subsection{Model-based Vorticity Compensation}\label{sec:f_hidden_model}

Similar to optical flow estimation,
it is challenging to estimate velocity from blurry observations.
When optimizing $\mathcal{L}_{\frac{D\sigma}{Dt}}$ with a blurry signal $\sigma$, a common issue is to have the rotational motion (referred to as vorticity) underestimated since velocity is only supervised along density gradient with $\mathbf{u} \cdot \nabla \sigma$.
Having $F_{vis}$ and $F_{hid}$ simultaneously optimized in our case, the two models have a chance to influence each other self-consistently through their blurry representation, resulting in sub-optimal solutions with underestimated vorticity.

To tackle this optimization difficulty,
we propose to use a model-based supervision for vorticity compensation.
The overall spatial distribution of the volume density has a strong correlation with the underlying motion, e.g., a smoke ring indicates strong and consistent vortices surrounding the annular region. %
Focusing on the relationship between density and velocity, \citet{chu2021learning} propose to train a GAN as a volume-to-volume translation network mapping a single density frame to its corresponding velocity frame. 
We denote their trained model as $d2v: \{\sigma\}_{n\times n \times n} \rightarrow \{\mathbf{u}\}_{n\times n \times n}$. We use this model as an additional supervision between the density volume generated by $F_{vis}$ and the velocity volume generated by $F_{hid}$. 

More specifically, we first sample $F_{vis}$ at uniform grid positions to get a density volume $\{d_{vis}\}_{32^3}$. Then $d2v$ model is used to generate a velocity reference $\{\mathbf{u}_{d2v}\}_{32^3} = d2v\Big(\{d_{vis}\}_{32^3}\Big)$.
Since $d2v$ model is trained in a relatively smaller data domain which cannot cover the variety of our fluid scenes, this velocity reference $\{\mathbf{u}_{d2v}\}_{32^3}$ can be different from the velocity volume $\{\mathbf{u}_{hid}\}_{32^3}$ sampled from the $F_{hid}$ in terms of absolute scale.
Thus, we normalize $\{\mathbf{u}_{d2v}\}_{32^3}$ and use it as a supervision on the vorticity distribution in the loss function:
\resizeEq{
\mathcal{L}_{d2v} & = \Bigg\|\frac{\nabla \times \mathbf{u}_{d2v}}{\sum_n^{32^3} \|\nabla \times \mathbf{u}_{d2v}\|_2^2 / 32^3 } - \frac{\nabla \times \mathbf{u}_{hid}}{\sum_n^{32^3} \|\nabla \times \mathbf{u}_{hid}\|_2^2  / 32^3}\Bigg\|_2^2 \ .
}{eq:d2vloss}{0.92\linewidth}
Note that $\nabla \times \mathbf{u}_{d2v}$ is calculated using numerical differences while $\nabla \times \mathbf{u}_{hid}$ is calculated with the automatic differentiation.
Jointly supervised with \myrefeq{eq:nseloss} and \myrefeq{eq:d2vloss}, our optimization can seek a more accurate solution with enhanced vorticity as well as reduced errors of $\mathcal{L}_{\frac{D\sigma}{Dt}}$ and $\mathcal{L}_{NSE}$, which will be demonstrated in the results.

\subsection{Tackling Color-Density Ambiguity}\label{sec:col-den}
NeRF~\cite{mildenhall2020nerf} and Neural Volumes~\cite{lombardi2019neural} require a high number of camera views to disentangle the color-density ambiguity.
When using sparse camera views, the color-density ambiguity is severe and over-fitting problems tend to occur. 
Across many scenes, we have observed a special form of over-fitting as ``ghost density'' artifact, which will be explained using an exemplar scene from ScalarFlow~\cite{eckert2019scalarflow} below.

\begin{figure}[tb] \footnotesize
\centering
\begin{overpic}[width=\linewidth]{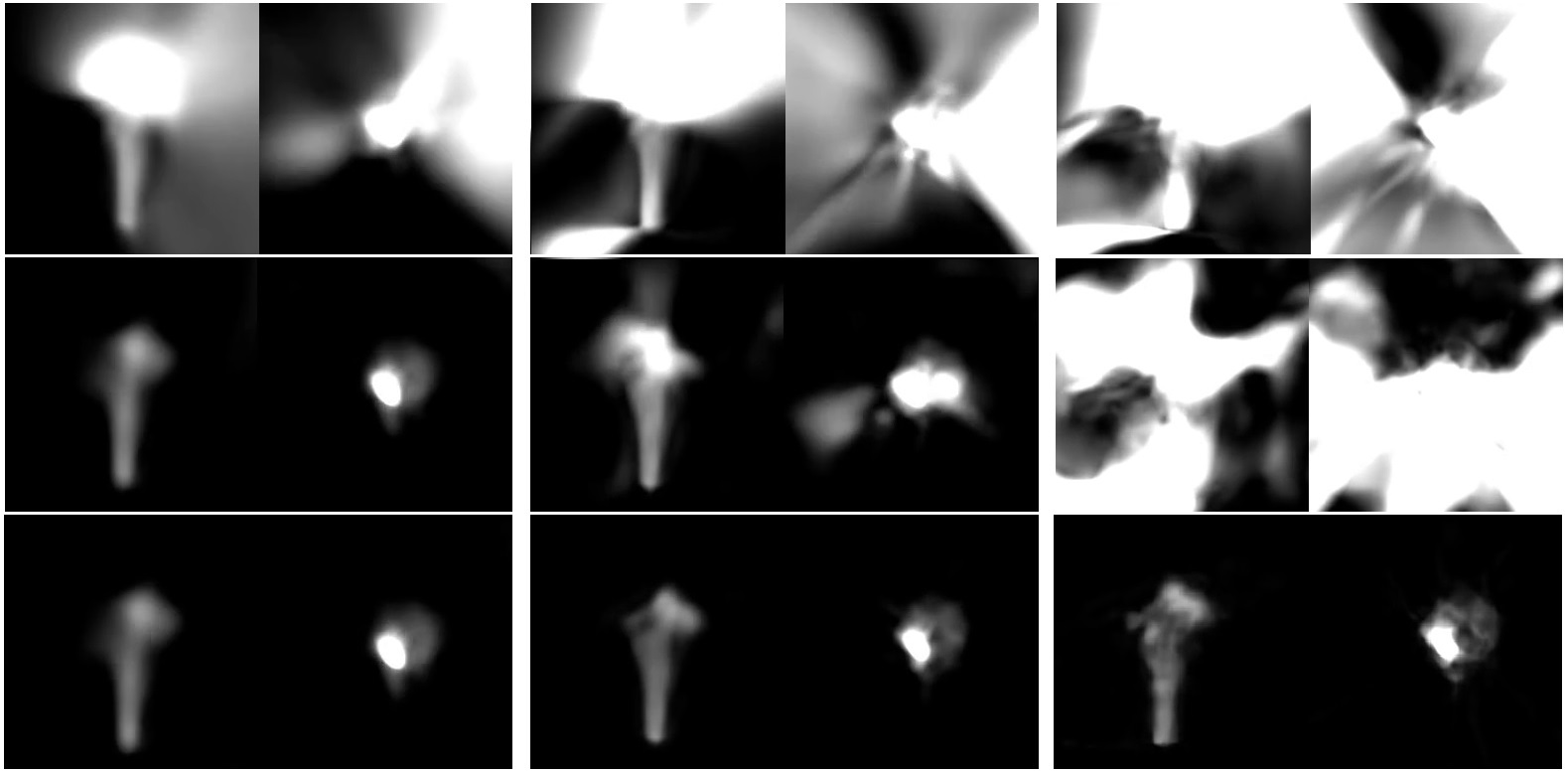}
\put(0,50){ 1k training iterations}
\put(33,50){ 5k training iterations}
\put(67,50){ 30k training iterations}
\put(-0.3,1.5){
\begin{tikzpicture}[x=0.1cm,y=0.1cm,z=0.1cm,>=stealth]
\draw[draw=white, ->] (xyz cs:x=0) -- (xyz cs:x=2.5) node[text=white,above] {$z$};
\draw[draw=white, ->] (xyz cs:y=0) -- (xyz cs:y=2.5) node[text=white,above] {$y$};
\end{tikzpicture}
}
\put(16.3,1.5){
\begin{tikzpicture}[x=0.1cm,y=0.1cm,z=0.1cm,>=stealth]
\draw[draw=white, ->] (xyz cs:x=0) -- (xyz cs:x=2.5) node[text=white,above] {$x$};
\draw[draw=white, ->] (xyz cs:y=0) -- (xyz cs:y=2.5) node[text=white,above] {$z$};
\end{tikzpicture}
}
\put(32.3,1.5){
\begin{tikzpicture}[x=0.1cm,y=0.1cm,z=0.1cm,>=stealth]
\draw[draw=white, ->] (xyz cs:x=0) -- (xyz cs:x=2.5) node[text=white,above] {$z$};
\draw[draw=white, ->] (xyz cs:y=0) -- (xyz cs:y=2.5) node[text=white,above] {$y$};
\end{tikzpicture}
}
\put(50.3,1.5){
\begin{tikzpicture}[x=0.1cm,y=0.1cm,z=0.1cm,>=stealth]
\draw[draw=white, ->] (xyz cs:x=0) -- (xyz cs:x=2.5) node[text=white,above] {$x$};
\draw[draw=white, ->] (xyz cs:y=0) -- (xyz cs:y=2.5) node[text=white,above] {$z$};
\end{tikzpicture}
}
\put(66.3,1.5){
\begin{tikzpicture}[x=0.1cm,y=0.1cm,z=0.1cm,>=stealth]
\draw[draw=white, ->] (xyz cs:x=0) -- (xyz cs:x=2.5) node[text=white,above] {$z$};
\draw[draw=white, ->] (xyz cs:y=0) -- (xyz cs:y=2.5) node[text=white,above] {$y$};
\end{tikzpicture}
}
\put(84.3,1.5){
\begin{tikzpicture}[x=0.1cm,y=0.1cm,z=0.1cm,>=stealth]
\draw[draw=white, ->] (xyz cs:x=0) -- (xyz cs:x=2.5) node[text=white,above] {$x$};
\draw[draw=white, ->] (xyz cs:y=0) -- (xyz cs:y=2.5) node[text=white,above] {$z$};
\end{tikzpicture}
}
\put(36.2,24.8){
\begin{tikzpicture}[x=0.1cm,y=0.1cm,z=0.1cm,>=stealth]
\filldraw[color=purple!50, fill=none, very thick](0,0) circle (2.5);
\end{tikzpicture}
}
\put(49.2,19.5){
\begin{tikzpicture}[x=0.1cm,y=0.1cm,z=0.1cm,>=stealth]
\filldraw[color=purple!50, fill=none, very thick](0,0) circle (2.5);
\end{tikzpicture}
}
\end{overpic}\\
a) Density profiles of \textit{SIREN+T} (top), \textit{Growing} (middle), \textit{Ours} (bottom) during training.
\\
\begin{overpic}[width=\linewidth]{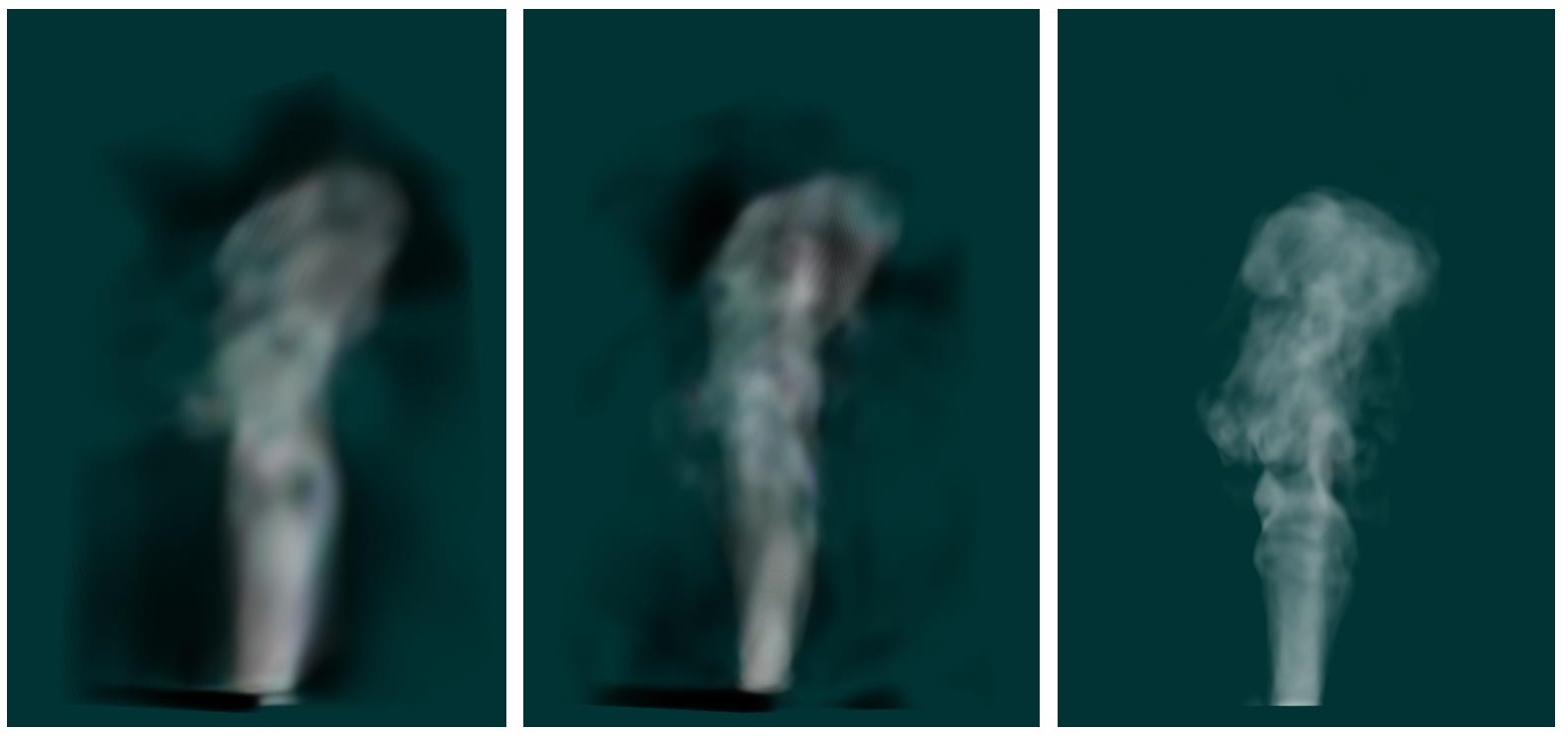}
\put(2,2){\color{white} b) \textit{SIREN+T}}
\put(35,2){\color{white} c) \textit{Growing}}
\put(69,2){\color{white} d) \textit{Ours}}
\end{overpic}
\caption{\footnotesize The ``ghost density'' artifact. 
a) Density profile (side and top views) during training. b),c), and d) Rendering results using a dark green background.}
\label{fig:ghost}
\end{figure}
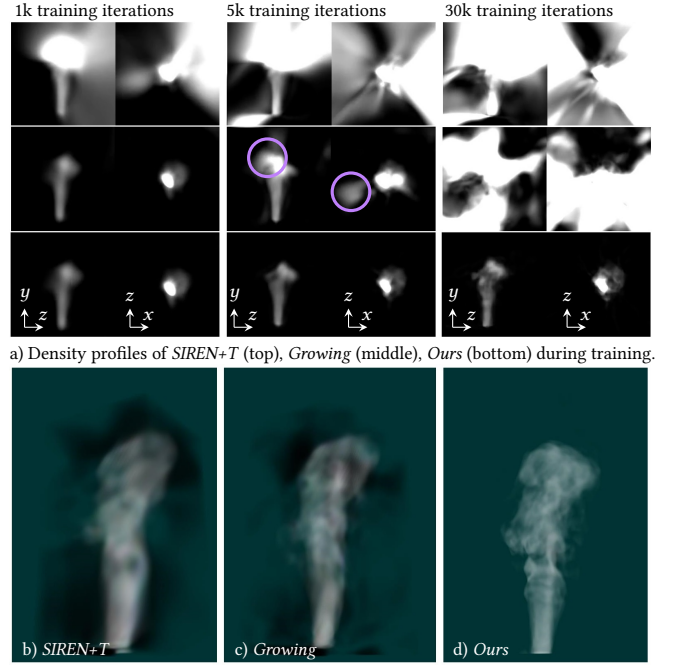

ScalarFlow Data are recorded from five cameras located on an 120 degree arc and the fluid volume is positioned in front of a black background.
Using the grey images captured from the five cameras, we train a time-varying NeRF model with SIREN layers (\textit{SIREN+T}) and its resulting density field is rendered as an RGBA image in \myreffig{fig:ghost}b with a dark green background. %
As shown in the image, the density of \textit{SIREN+T} fills most of the 3D region. With this density region acting as a ``canvas'', \textit{SIREN+T} model simply learns to ``paint'' appropriate color to match training views. While it manages to fulfill the rendering objective function $\mathcal{L}_{\widetilde{img}}$, it produces a radiance field far from the ground-truth with a lot of density in the background. We denote the artifact as ``ghost density'', which represents a typical failure to disentangle the color-density ambiguity.

To tackle the color-density ambiguity, we propose to use a progressively growing model to alleviate overfitting and use a regularization term to penalize the ``ghost density''.
In line with the coarse-to-fine optimization proposed by \citet{park2020nerfies} for coordinate-based MLPs with positional encoding, e.g. the original NeRF model, we propose a layer-by-layer growing strategy for the MLPs without positional encoding including the SIREN-based~\cite{sitzmann2020implicit} models. 
During training, 
we use a sliding window to first select neurons from early layers like training a shallow model,
then gradually slide the window towards the following layers until the full architecture is covered.
On each hidden layer $m\in [0,N-1]$, the sliding window is represented by the weighting parameter 
$w_{lm} =  clamp(1+m_a-m,0,1)\cdot clamp(1+m-m_a,0,1)$, with $ m_a = 1 + (N-2) s/S$ in the range from $1$ to $N-1$ which semantically represents the current last hidden layer, $s$ stands for the current training iteration step, and $S$ stands for the iteration step when the progressive growing accomplishes. As a further explanation, layer $m=0$ is a permanent hidden layer with $w_{l0} = 0$, layer $m=1$ only fades out with $w_{l1} = clamp(1+m-m_a,0,1)$, intermediate layers fade in and out, and the last layer $m=N-1$ only fades in. 

From \myreffig{fig:ghost}a, we can see that the \textit{SIREN+T} model in the first row has density spreading across the domain at the beginning of the training.
After applying the progressive growing, the \textit{Growing} model in the second row learns a reasonable density profile at the beginning, however, "ghost density" occurs at the iteration step 5k, marked in purple circles.
To penalize the ``ghost density'', we propose a regularization in 2D image space:
\resizeEq{
\mathcal{L}_{ghost}\Big(\mathbf{C}(r_{ijt}),\mathbf{B}_{ijt}, a(r_{ijt}) \Big) = {\tt sigmoid}\Big[-\Big(\mathbf{C}(r_{ijt})-\mathbf{B}_{ijt}\Big)^2\Big]\cdot A(r_{ijt}) \ .
}{eq:ghostloss}{0.92\linewidth}
Taking the background color $\mathbf{B}_{ijt}$ as a known input, $\mathcal{L}_{ghost}$ penalizes the opacity $A(r_{ijt})=  \sum_{k=1}^{K} T(h_k) \alpha(h_k)$ in 2D image space when the rendered pixel color $\mathbf{C}(r_{ijt})$ is close to the background color.
Applying $\mathcal{L}_{ghost}$ on the progressively growing model, the third row in \myreffig{fig:ghost}a properly refines the density profile during training.
From c), d), and e) in \myreffig{fig:ghost}, the \textit{Growing} model reduces ``ghost density'' and captures more details on the volume than the \textit{SIREN+T} model, while our full model achieves the best result without ``ghost density''.
Our approach to removing the ``ghost density'' artifact is effective across many scenes, which will be shown in the result section. All testing cases use a background of a solid color like dark green, which can be seen as being "transparent".

\subsection{Extensions for hybrid scenes with static obstacles}\label{sec:hybrid}
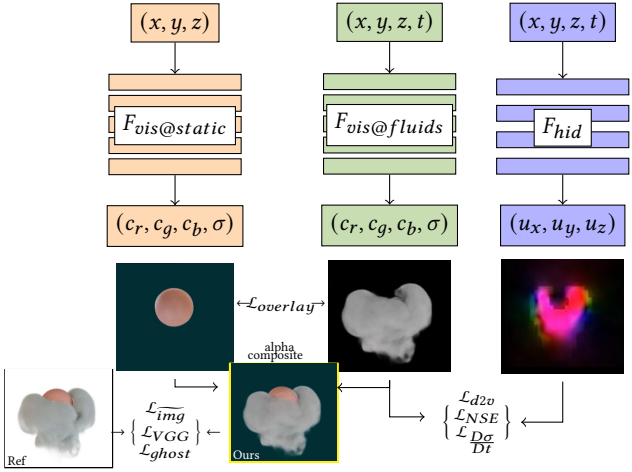
\begin{figure}[thb]
    \begin{tikzpicture}
    \node (in1) [draw=black,fill=orange!30] {$(x,y,z)$};
    
    \node (pro1) [below of=in1, yshift=6pt, draw=black, fill=orange!30, minimum width=50pt, minimum height=6pt] { };
    \node (pro2) [below of=in1, yshift=-2pt, draw=black, fill=orange!30, minimum width=50pt, minimum height=6pt] { };
    \node (pro3) [below of=in1, yshift=-10pt, draw=black, fill=orange!30, minimum width=50pt, minimum height=6pt] { };
    \node (pro4) [below of=in1, yshift=-18pt, draw=black, fill=orange!30, minimum width=50pt, minimum height=6pt] { };
    \node (pro5) [below of=in1, yshift=-26pt, draw=black, fill=orange!30, minimum width=50pt, minimum height=6pt] { };
    
    \node (staticM) [below of=pro1, yshift=12pt, draw=black, fill=white!30] { $F_{vis@static}$ };
    
    \node (output1) [below of=pro5, yshift=6pt, draw=black, fill=orange!30] {$(c_r,c_g,c_b,\sigma)$};
    
    \node (fig1) [below of=output1, yshift=-6pt] {\includegraphics[width=44pt]{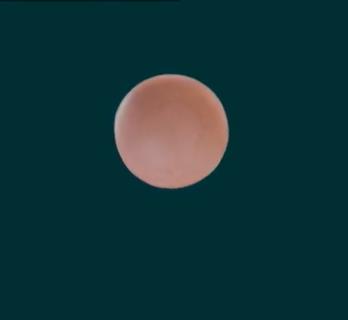}
    };
    \node (margin) [below of=fig1, yshift=2pt,minimum height=6pt,minimum width=32pt] {~}; %
    \node (figO) [below of=fig1, draw=black,  line width=0.6pt, inner sep=0.3pt, yshift=-10pt, xshift=-44pt] 
    { \begin{overpic}[width=40pt]{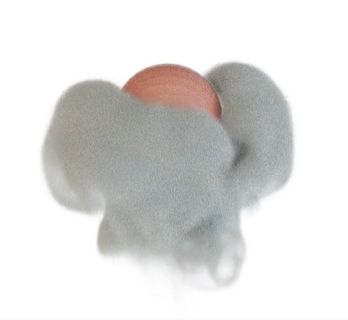}
        \put(2,2){\tiny Ref}
    \end{overpic}};
    
    \node (eq1) [right of=figO, xshift= 12pt, yshift=-5pt] {\tiny  $ 
    \rightarrow \Bigl\{\begin{smallmatrix}
       \mathcal{L}_{\widetilde{img}}\\ 
       \mathcal{L}_{VGG} \\
       \mathcal{L}_{ghost} \\
     \end{smallmatrix}\Bigr\} \leftarrow$};

    \draw [->] (in1) -- (pro1);
    \draw [->] (pro5) -- (output1);
    \draw [->] (fig1) |- node[left,below]{~}(margin.east);
    \end{tikzpicture}
    \hspace{-10pt}
    \begin{tikzpicture}
    \node (in2) [draw=black, fill=green!30] {$(x,y,z,t)$};
    
    \node (spro1) [below of=in2, yshift=6pt, draw=black, fill=green!30, minimum width=50pt, minimum height=6pt] { };
    \node (spro2) [below of=in2, yshift=-2pt, draw=black, fill=green!30, minimum width=50pt, minimum height=6pt] { };
    \node (spro3) [below of=in2, yshift=-10pt, draw=black, fill=green!30, minimum width=50pt, minimum height=6pt] { };
    \node (spro4) [below of=in2, yshift=-18pt, draw=black, fill=green!30, minimum width=50pt, minimum height=6pt] { };
    \node (spro5) [below of=in2, yshift=-26pt, draw=black, fill=green!30, minimum width=50pt, minimum height=6pt] { };
    
    \node (smokeM) [below of=spro1, yshift=12pt, draw=black, fill=white!30] { $F_{vis@fluids}$ };
    
    \node (output2) [below of=spro5, yshift=6pt, draw=black, fill=green!30] {$(c_r,c_g,c_b,\sigma)$};
    \node (fig2) [below of=output1, yshift=-6pt] {
    \begin{overpic}[width=46pt]{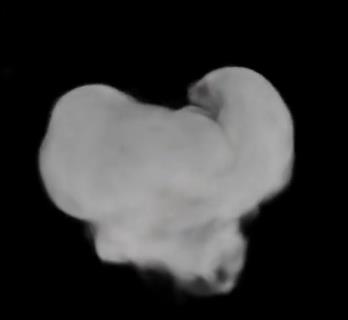}
    \end{overpic}};
    \node (figO) [below of=fig2, draw=yellow,  line width=0.8pt, inner sep=0.4pt, yshift=-8pt, xshift=-40pt]{
    \begin{overpic}[width=40pt]{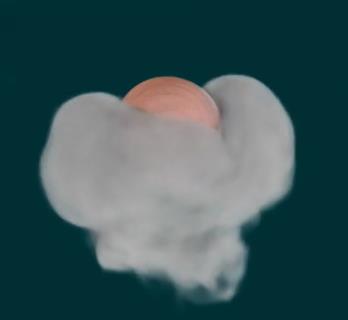}
        \put(2,2){\color{white} \tiny Ours}
        \put(32,104){\color{black} \tiny alpha}
        \put(16,96){\color{black} \tiny composite}
        \put(5,146){{\tiny $\leftarrow$\hspace{-2pt}$\mathcal{L}_{overlay}$\hspace{-2pt}$\rightarrow$}}
    \end{overpic}};
    \node (margin2) [below of=figO, yshift=38pt, minimum width=40pt] {~};
    \node (margin) [right of=fig2, xshift=-12pt, yshift=-38pt, minimum height=40pt] {~};
   
    \draw [->] (in2) -- (spro1);
    \draw [->] (spro5) -- (output2);
    \draw [->] (fig2) |- node[right,below]{~}(margin2);
    \draw [->] (fig2) |- node[left,below]{~}(margin);
    \end{tikzpicture}
    \hspace{-14pt}
    \begin{tikzpicture}
    \node (in2) [draw=black, fill=blue!30] {$(x,y,z,t)$};
    
    \node (spro1) [below of=in2, yshift=4pt, draw=black, fill=blue!30, minimum width=50pt, minimum height=6pt] { };
    \node (spro2) [below of=in2, yshift=-6pt, draw=black, fill=blue!30, minimum width=50pt, minimum height=6pt] { };
    \node (spro3) [below of=in2, yshift=-16pt, draw=black, fill=blue!30, minimum width=50pt, minimum height=6pt] { };
    \node (spro4) [below of=in2, yshift=-26pt, draw=black, fill=blue!30, minimum width=50pt, minimum height=6pt] { };
    
    \node (smokeM) [below of=spro1, yshift=13pt, draw=black, fill=white!30] { $F_{hid}$ };
    
    \node (output2) [below of=spro4, yshift=6pt, draw=black, fill=blue!30] {$(u_x,u_y,u_z)$};
    
    \node (fig2) [below of=output1, yshift=-6pt] {\includegraphics[width=46pt]{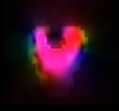}};
    
    \node (eq2) [left of=fig2, xshift= -4pt, yshift= -40pt, minimum height=36pt] {\footnotesize
    $\Bigl\{\begin{smallmatrix}
        \mathcal{L}_{d2v}\\
       \mathcal{L}_{NSE} \\
        \mathcal{L}_{\frac{D\sigma}{Dt}}
     \end{smallmatrix}\Bigr\}
    $};
    \draw [->] (fig2) |- node[right,below]{~}(eq2); 
    \draw [->] (in2) -- (spro1);
    \draw [->] (spro4) -- (output2);
    \end{tikzpicture}
\caption{\footnotesize We use hybrid Models to learn radiance fields for static obstacles and dynamic fluids.
The velocity model is only related to the density of fluids.}
\label{fig:structure}
\end{figure}

In addition to the fluid part addressed above, we extend the NeRF-based method to handle hybrid scenes of dynamic fluid with static obstacles. Specifically, we design a variant of the proposed architecture to separate the dynamic and the static parts in an unsupervised manner.
As shown in \myreffig{fig:structure}, our hybrid $F_{vis}$ model consists of two sub-models for a scene. One model, $F_{vis@static}$, represents the static obstacle with 3D positions $(x,y,z)$ as input and the other one, $F_{vis@fluids}$, handles time-varying radiance with a 4D input $(x,y,z,t)$.
Since $F_{vis@fluids}$ can represent the static part as well, we defer the training of this model. That said, $F_{vis@static}$ is trained first to describe the whole scene as much as possible and the time-varying dynamic part can be taken over by $F_{vis@fluids}$ gradually.

For particular scenes, additional constraints are helpful in practice, e.g., applying a time-varying hull when tracing $F_{vis@fluids}$, adding a hue constraint on $F_{vis@fluids}$, or capturing more static images for the scene without fluids as references for $F_{vis@static}$. Instead, our method solely based on the hybrid architecture and the sequential training strategy. This allows us to handle general scenes and achieves an unsupervised separation of obstacles and fluids.

To train the new $F_{vis}$ and a single $F_{hid}$, the objective functions are adjusted accordingly. See \myreffig{fig:structure} for a summary illustration of the objective functions. The supervision in 2D image space, i.e. $\mathcal{L}_{img} + w_{VGG}\cdot\mathcal{L}_{VGG} + w_{ghost}\cdot\mathcal{L}_{ghost^*}$, needs the objective color $\mathbf{C}_{compos}(r_{ijt})$ calculated from the alpha composite of the static and dynamic components:
\resizeEq{
\mathbf{C}_{compos}(r_{ijt}) = &  \sum_{k=1}^{K} T(h_k) \Big( \sum_{model}^{static,fluids} \alpha_{model}(h_k) \mathbf{c}_{model}(h_k) \Big), \\
{A}_{compos}(r_{ijt}) = &  \sum_{k=1}^{K} T(h_k) \Big( \alpha_{static}(h_k)  + \alpha_{fluids}(h_k)  \Big),
}{eq:color_compos}{0.89\linewidth}
where $T(h_k) = exp\Big(-\sum_{\hat{k}=1}^{k-1} \sigma_{static}(h_{\hat{k}}) + \sigma_{fluids}(h_{\hat{k}}) \delta_{\hat{k}} \Big)$.
We merely adjust the ghost density regularization as:
\resizeEq{
\mathcal{L}_{ghost^*} = & \mathcal{L}_{ghost}\Big(\mathbf{C}_{compos}(r_{ijt}), \mathbf{B}_{ijt}, A_{compos}(r_{ijt})\Big) \\
& + \mathcal{L}_{ghost}\Big( \mathbf{C}_{static}(r_{ijt}), \mathbf{B}_{ijt}, A_{static}(r_{ijt})\Big) \\
& + \mathcal{L}_{ghost}\Big( \mathbf{C}_{fluids}(r_{ijt}), \mathbf{C}_{static}(r_{ijt}), A_{fluids}(r_{ijt})\Big) \ .
}{eq:ghostlosshyb}{0.8\linewidth}
On the other hand, the spatio-temporal motion supervision of $\mathcal{L}_{\frac{D\sigma}{Dt}} + w_{NSE}\mathcal{L}_{NSE} + w_{d2v}\mathcal{L}_{d2v}$ is only related to the physics-informed velocity model $F_{hid}$ and the dynamic sub-model $F_{vis@fluids}$. 
Additionally, a volumetric overlay loss, $\mathcal{L}_{overlay} = \frac{\sigma_{static}\cdot \sigma_{fluids}}{\sigma_{static}^2 + \sigma_{fluids}^2}$, is proposed to reduce the intersection of static and dynamic components, since fluids should not exist inside obstacles.

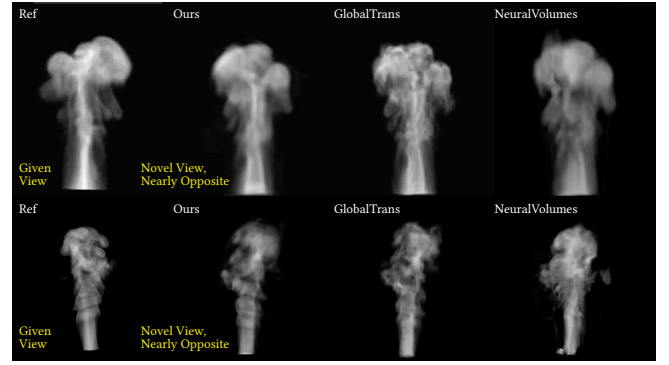
\begin{figure}[tb] \tiny
    \centering
    \begin{overpic}[width=\linewidth]{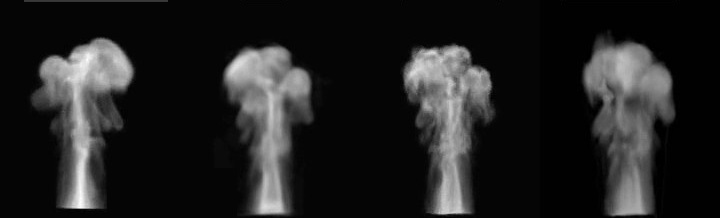}
    \put(1,28){\color{white} Ref}
    \put(1,4){{\color{yellow} Given}}
    \put(1,2){{\color{yellow} View}}
    \put(25,28){\color{white} Ours}
    \put(50,28){\color{white} GlobalTrans}
    \put(75,28){\color{white} NeuralVolumes}
    \put(20, 4){\color{yellow} Novel View,}
    \put(20, 2){\color{yellow} Nearly Opposite}
    \end{overpic}\\
    \begin{overpic}[width=\linewidth]{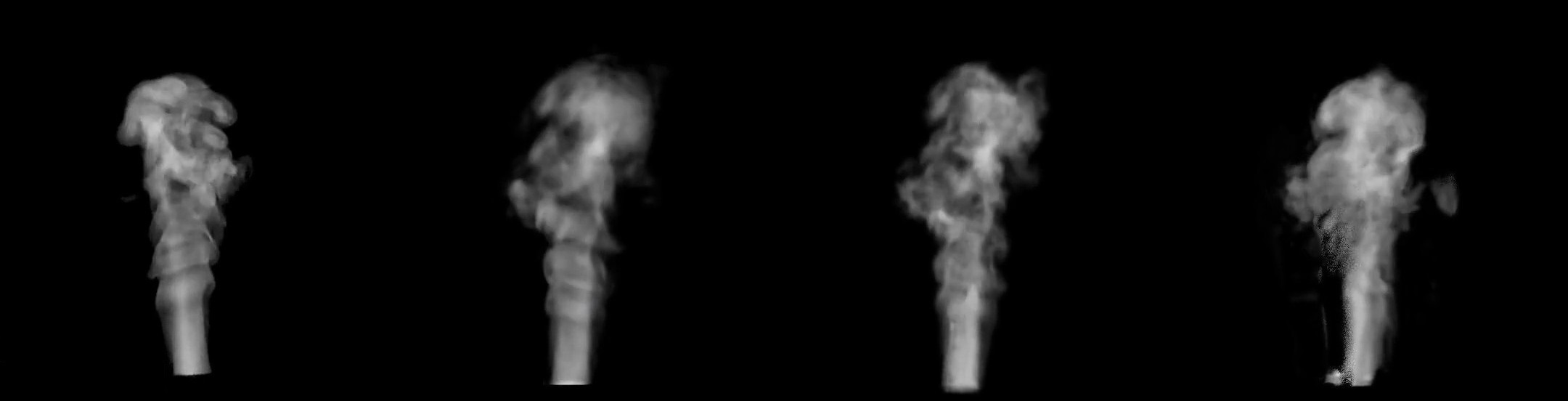}
    \put(1,23){\color{white} Ref}
    \put(1,4){{\color{yellow} Given}}
    \put(1,2){{\color{yellow} View}}
    \put(25,23){\color{white} Ours}
    \put(50,23){\color{white} GlobalTrans}
    \put(75,23){\color{white} NeuralVolumes}
    \put(20,4){\color{yellow} Novel View,}
    \put(20,2){\color{yellow} Nearly Opposite}
    \end{overpic}\\
    \caption{Comparisons of rendering results on the ScalarFlow Dataset with synthetic (1st row) and real (2nd row) cases.}
    \label{fig:scalar}
\end{figure}

\section{Results and Evaluation}\label{sec:evaluation}

We test our algorithm on synthetic and real fluid scenes with a wide variety.
\myrefsec{sec:ScalarTest} shows qualitative and quantitative results on the ScalarFlow dataset~\cite{eckert2019scalarflow} with synthetic and real captured plume.
Synthetic scenes in \myrefsec{sec:SynthticTest} are rendered using Blender %
to test regular fluid settings under complex lighting conditions.
At last, hybrid scenes of fluid flow with complex obstacles are tested in \myrefsec{sec:ObsTest}.
We refer the readers to our supplemental material as a webpage with corresponding video clips that more clearly display the quality of the reconstructed radiance and motion fields.

\subsection{ScalarFlow Dataset}\label{sec:ScalarTest}

As mentioned above, ScalarFlow dataset captures recordings of a real plume using five fixed cameras located on a 120-degree arc. After post-processing steps, like the subtraction of the first, empty frame, the ScalarFlow recordings have a clean background in black. In addition, ScalarFlow dataset has synthetic data of a virtual plume simulated using a numerical fluid solver~\cite{mantaflow2018}.

{\footnotesize
\setlength{\tabcolsep}{3pt}
\setlength\extrarowheight{1pt}
\begin{figure}[tbh]
\begin{minipage}[t]{\linewidth}
    \centering
    \includegraphics[width=\linewidth]{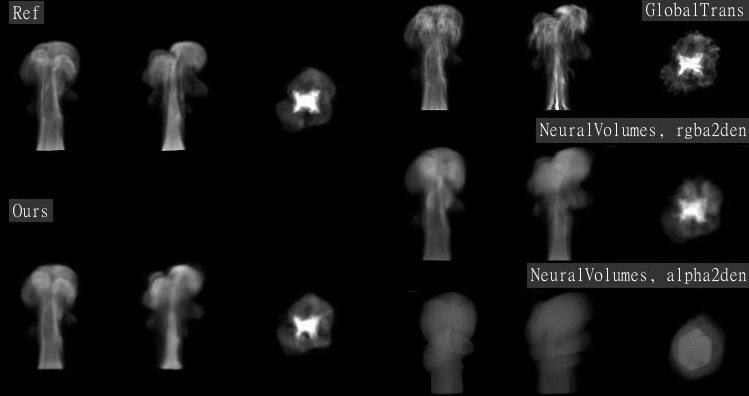}\\
    \includegraphics[width=\linewidth]{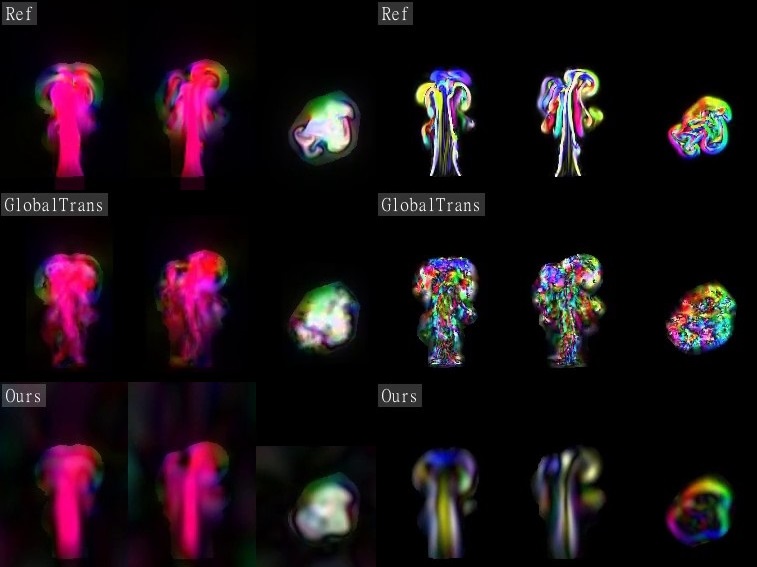}
    \caption{Reconstructed volume of density (top), velocity (bottom left) and vorticity (bottom right) for the synthetic ScalarFlow data. We show the front, side, and top views. Velocity and vorticity are visualized using the middle slice.
    Velocity inside the visual hull is visualized at full intensity, and reduced intensity ($\times0.3$) is used outside.}
    \label{fig:scalar_syn1}
\end{minipage}
\begin{minipage}[t]{\linewidth}
\captionof{table}{{Comparisons with related work on the synthetic ScalarFlow Data.}}\label{tab:scalar}
\centering
\begin{tabular}{l|c|c|c|c|c} 
\hline
$ ^{\text{Volumetric}}_{\text{Evaluation}}$
& $l_2(\sigma, \sigma_{ref}) \ \downarrow$  
& $l_2(\mathbf{u}, \mathbf{u}_{ref}) \ \downarrow$  
& $\nabla \cdot \mathbf{u} \ \downarrow$
& {\tiny $^{\text{Warp}}_{\text{Error}} \ \downarrow$ }      
& {\tiny $^{\text{MidWarp}}_{\text{Error}} \ \downarrow$ }     \\[2pt] \hline
\textit{Reference} &
    \textit{0}        & \textit{0}        & \textit{0.0339} & \textit{0.3212}  & \textit{0.0664} \\[2pt] \hline
$ ^{\text{Neural}}_{\text{Volumes}}$    &  $^{\text{5.06 (rgba2den) }}_{\text{11.20 (alpha2den) }}$  & - & - & - & -   \\[4pt] \hline
$ ^{\text{Global}}_{\text{Transport}}$         &
    3.9324           & 0.5073          & 0.0592          & \textbf{0.0917} & 0.3905           \\[2pt] \hline
Ours               &
   \textbf{3.6264}          & \textbf{0.4475} & \textbf{0.0353} & 0.2371          & \textbf{0.1581} \\[2pt] \hline
\end{tabular}
\end{minipage}
\end{figure}
}

\paragraph{Synthetic Data}

In the test on the synthetic data, a regular fluid flow is generated in resolution of $128\times192\times128$. 
We simulate 120 steps with a time step of 0.5 ($0\leq t \leq 60$).
The first 60 steps ($0\leq t \leq 30$) are skipped as a startup and
we use every second of the remaining 60 steps ($t = \{30, 31, ..., 59\}_{30}$)
as the target fluid fields to be reconstructed.
For consistent comparison, the differentiable rendering of Global Transport~\cite{franz2021global} is used to render images at given time steps with a empirically determined lighting condition that roughly matches the lighting of the real captures, and black color is used as the background.

\revise{
In the following, we first evaluate the proposed method (\textit{Ours})
against related work including the Global Transport method~\cite{franz2021global} (\textit{GlobalTrans}) and Neural Volumes~\cite{lombardi2019neural} (\textit{NeuralVolumes}).
Then, to illustrate the role of each term in our supervision, ablation studies are presented with our full model and three ablated ones.
We provide qualitative evaluations via rendered images and video clips (Sec. 1.1 of the supplemental webpage). 
Quantitative comparisons are given with the numerical errors on the reconstructed volumetric attributes (\myreftab{tab:scalar},~\ref{tab:scalar1}).
}

In the second and third columns of \myreftab{tab:scalar}, we calculate the averaged $l_2$ errors, $l_2(\sigma, \sigma_{ref}) = \|\sigma - \sigma_{ref}\|_2^2$ and $l_2(\mathbf{u}, \mathbf{u}_{ref}) = \|\mathbf{u} - \mathbf{u}_{ref}\|_2^2$, on the reconstructed density and velocity volume with regard to the reference. 
Note that the reference and results from previous work have a discretized representation, thus our results are sampled from our continuous models at uniform grid positions to compare consistently using numerical metrics.
While our models and \textit{NeuralVolumes} simply apply a 3D bounding box as the domain of the grid in $128\times192\times128$, \textit{GlobalTrans} performs optimization inside a time-varying visual hull which is projected from the 2D smoke region in the given images.
All metrics are measured inside this visual hull with the inflow region (the bottom $128\times20\times128$ grid cells) excluded.
Corresponding to the visualizations shown in \myreffig{fig:scalar_syn1}, the density reconstructed by \textit{Ours} and \textit{GlobalTrans} achieves similar accuracy, but \textit{GlobalTrans} exhibits high-frequency noise not present in the reference. However, the density reconstructed from the opacity of \textit{NeuralVolumes} (denoted as "alpha2den") is rather uniform, leading to an $l_2$ error of $11.2$.
In addition, we generate another density via $0.01\alpha(c_r+c_g+c_b)$ denoted as "rgba2den", which removes the ``ghost density'' and achieves an $l_2$ error of $5.06$. This indicates that \textit{NeuralVolumes} fails in the density-color disentanglement. While \textit{GlobalTrans} has density and color disentangled based on the given lighting condition, our method
manages to disentangle properly with the proposed growing strategy and the regularization term $\mathcal{L}_{ghost}$ without knowing the lighting condition.

Velocity and vorticity visualizations use the middle slices from front, side, and top views. For all methods, we reduced the velocity intensity outside the visual hull for visualization, since the region with non-zero density is more important.
\textit{NeuralVolumes} does not provide a velocity field. Its results also exhibit discontinuity in time. 
\textit{GlobalTrans} achieves better accuracy on the velocity but introduces high-frequency noise, which is also more visible in its vorticity field.
The velocity of our method has the minimal $l_2$ error, smallest divergence, and small warping errors, as shown in the last three columns of the table. 
\begin{figure}
{\footnotesize
Warp-Error \hfill MidWarp-Error \\
$\|{\tt Adv}(\sigma_t, \mathbf{u}_t) - \sigma_{t+1}\|_2^2 $ \hfill
$\|{\tt Adv}(\sigma_{t+1}, -0.5\mathbf{u}_{t+1}) \  -{\tt Adv}(\sigma_t, 0.5\mathbf{u}_t) \|_2^2$ }\\

\begin{overpic}[width=0.9\linewidth]{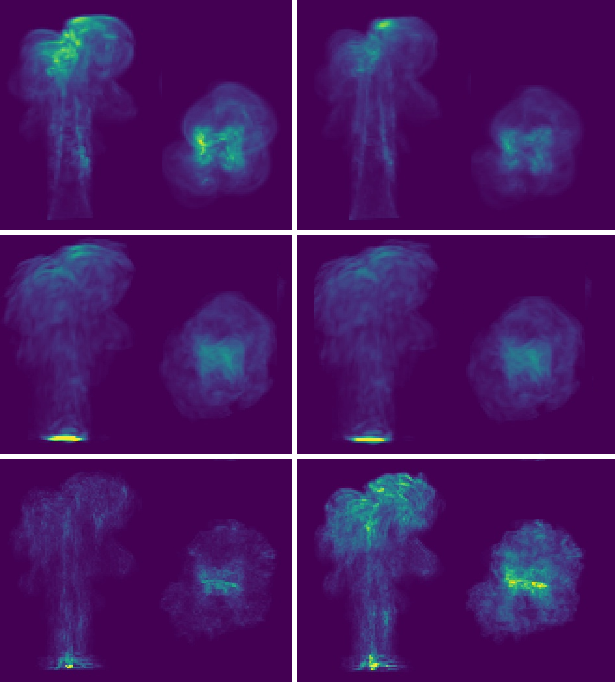}
\put(62,2){\color{white} GlobalTrans}
\put(62,35){\color{white} Ours}
\put(62,68){\color{white} Reference}
\put(22,2){\color{white} GlobalTrans}
\put(22,35){\color{white} Ours}
\put(22,68){\color{white} Reference}
\end{overpic}
    \caption{\footnotesize Visualization of the warping error measured using two metrics. The left one uses a full step forward warping, and the right one uses a half step forward warping and a half step backward warping.
    We show the side and top views.}
    \label{fig:scalar_syn2}%
\end{figure}

We compare the warping error using two metrics in \myreffig{fig:scalar_syn2}. The first metric $\|{\tt Adv}(\sigma_t, \mathbf{u}_t) - \sigma_{t+1}\|_2^2 $, denoted as "Warp-Error", represents a frame to frame warping.
Denoted as "MidWarp-Error", the second metric $\|{\tt Adv}(\sigma_{t+1}, -0.5\mathbf{u}_{t+1}) \  -{\tt Adv}(\sigma_t, 0.5\mathbf{u}_t) \|_2^2$ shows the error between two consecutive density frames advected to the midpoint time using forward and backward warping, respectively.
Note that all reconstruction models see images at time $t$ and $t+1$, but the frames at the midpoint of $t+0.5$ are not given. The reference is numerically simulated with a time step of 0.5.
\textit{GlobalTrans} has the minimal frame-to-frame warping error
which is even smaller than the reference, since it is trained to minimize the transport error at this discrete level.
As a continuous method, \textit{Ours} has a smaller midpoint warping error than the full step warping error, which is in consistent with the reference, and \textit{Ours} has the smallest "MidWarp-Error".
While it is rigorous to evaluate the warping metric at an unseen midpoint time, it is the situation %
in reality where continuous fluid phenomena are captured by cameras with limited frame-rates.

{\footnotesize
\setlength{\tabcolsep}{3pt}
\setlength\extrarowheight{1pt}
\begin{figure}[bth]
\begin{minipage}[t]{\linewidth}
    \centering
    \includegraphics[width=0.9\linewidth]{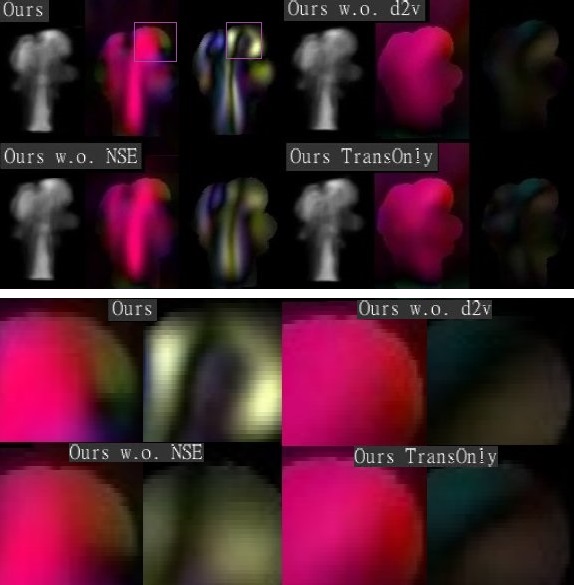}
    \caption{\revise{1st row: Ablation study with reconstructed volume of density (left), velocity (middle), and vorticity (right) for the synthetic ScalarFlow data. 2nd row: Velocity (left) and vorticity (right) zoomed in, corresponding to the purple rectangles on the 1st row.}}
    \label{fig:scalar_syn3}
\end{minipage}
\begin{minipage}[t]{\linewidth}
\captionof{table}{\revise{Ablation study on the synthetic ScalarFlow Data.}}\label{tab:scalar1}
\centering
\begin{tabular}{l|c|c|c|c|c} 
\hline
$ ^{\text{Volumetric}}_{\text{Evaluation}}$
& $l_2(\sigma, \sigma_{ref}) \ \downarrow$  
& $l_2(\mathbf{u}, \mathbf{u}_{ref}) \ \downarrow$  
& $\nabla \cdot \mathbf{u} \ \downarrow$
& {\tiny $^{\text{Warp}}_{\text{Error}} \ \downarrow$ }      
& {\tiny $^{\text{MidWarp}}_{\text{Error}} \ \downarrow$ }     \\[2pt] \hline
\textit{Ours}               &
   \textit{3.6264}          & \textbf{\textit{0.4475}} & \textbf{\textit{0.0353}} & \textit{0.2371 }        & \textit{0.1581} \\[2pt] \hline
$ ^{\text{Ours}}_{\text{w.o. d2v}}$     &
    \textbf{3.4926}          & 0.8786          & 0.0653          & 0.2359          & 0.2192         \\[2pt] \hline
$ ^{\text{Ours}}_{\text{w.o. NSE}}$     &
    3.8613          & 0.4883          & 0.0427          & 0.0905          & 0.0786         \\[2pt] \hline
$ ^{\text{Ours}}_{\text{TransOnly}}$     &
    3.5060          & 0.9127          & 0.0689          & \textbf{0.0466}          & \textbf{0.0402}         \\[2pt] \hline
\end{tabular}
\end{minipage}
\end{figure}
}
\revise{
We further conduct ablation studies on the synthetic scene using the same evaluation metrics.
In the first ablated model (\textit{Ours w.o. d2v}), we remove $\mathcal{L}_{d2v}$, the term for vorticity compensation.
The second ablated model (\textit{Ours w.o. NSE}) has dropped the $\mathcal{L}_{NSE}$ term to show the role of physical priors.
In the last ablated model (\textit{Ours TransOnly}), we remove both $\mathcal{L}_{d2v}$ and $\mathcal{L}_{NSE}$,
so that its velocity field is only supervised with the transport equation $\mathcal{L}_{\frac{D\sigma}{Dt}}$.
The supervision applied on the radiance fields is not changed.

As shown in the first column of \myreftab{tab:scalar1},
our full and ablated models behave similarly on the density reconstruction, with an average $l_2$ error of 3.6212 and their differences in the range of $\pm 6\%$.
While \textit{Ours TransOnly} has the smallest warping errors due to its velocity field only supervised by the transport equation, its velocity  actually differs the most from the ground truth with an $l_2$ error of 0.9127.
With our comprehensive supervision, \textit{Ours} manages to reconstruct velocity more accurately.
By comparing \textit{Ours w.o. d2v} to \textit{Ours} and comparing \textit{Ours TransOnly} to \textit{Ours w.o. NSE}, we see consistently that the former models without $\mathcal{L}_{d2v}$ can only roughly match the reference velocity and have vorticity fields that are significantly underestimated.
Merely using differentiable equations as objective functions, their nonlinearity makes the optimization process of the former models difficult. With the help of $\mathcal{L}_{d2v}$ as model-based supervision, the latter models present enhanced vorticity and are much closer to the reference.
Meanwhile, when comparing \textit{Ours w.o. NSE} to \textit{Ours} and comparing \textit{Ours TransOnly} to \textit{Ours w.o. d2v}, Navier-Stokes equations serve as physical priors and help in generating physically plausible but not necessarily correct solutions. 
We see that the latter models always have smaller divergence compared to the former ones.
However, the result of \textit{Ours w.o. d2v} shows little improvement over \textit{Ours TransOnly} in \myreffig{fig:scalar_syn3}, both suffering from vorticity underestimation.
When using our full supervision, Navier-Stokes equations contribute more to the correct solution.
Compared to \textit{Ours w.o. NSE}, \textit{Ours} presents sharper vorticity and has more details in the velocity fields.
}

\begin{figure}[tbh]\footnotesize
    \centering
    \includegraphics[width=\linewidth]{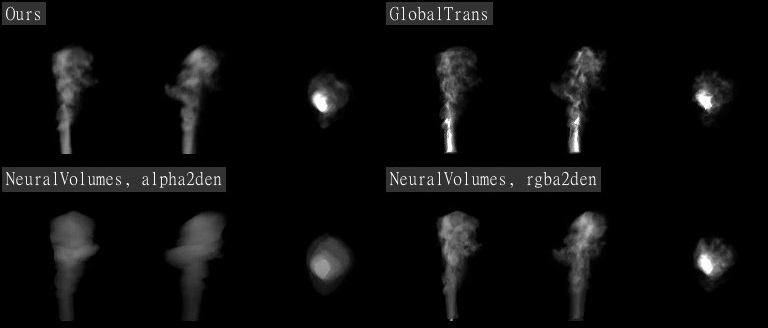}\\
    \includegraphics[width=\linewidth]{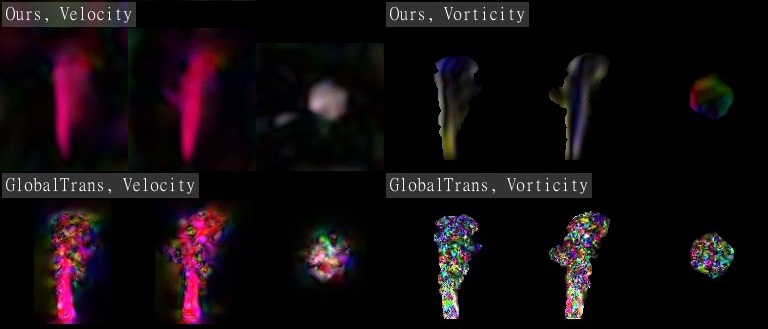}\\
    { Warp-Error \hfill MidWarp-Error} \\
    \begin{overpic}[width=\linewidth]{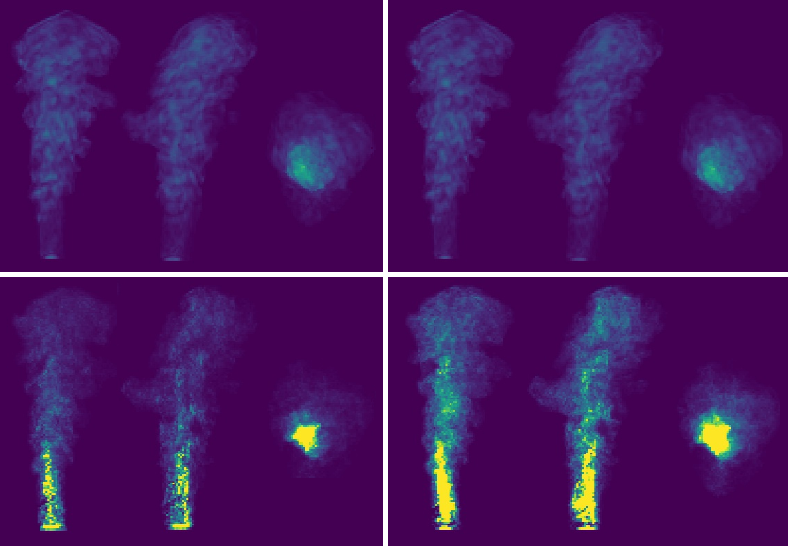}
    \put(82,1){\color{white} GlobalTrans}
    \put(90,36){\color{white} Ours}
    \put(32,1){\color{white} GlobalTrans}
    \put(40,36){\color{white} Ours}
    \end{overpic}
    \caption{Reconstructed volume of density (top), velocity (middle left) and vorticity (middle right) for the real captures of ScalarFlow. Again,
    velocity visualization is reduced outside the visual hull.
    Metrics evaluating warping error with a full time step and with half steps are shown at bottom.}
    \label{fig:scalar_real1}
\end{figure}

\paragraph{Real Captures}
The real fluid captures of ScalarFlow have a frame-rate of 60. We take the middle 120 frames from each camera view for fluid reconstruction.
With similar settings on camera calibration and lighting conditions,
the conclusions for the real captures are consistent with the synthetic case.
\textit{NeuralVolumes} trivially encodes density information into color fields,
as shown in \myreffig{fig:scalar_real1}.
With given lighting conditions and visual hull applied, \textit{GlobalTrans} generates detailed density and velocity fields with high-frequency noise. 
While our density volume is not as detailed as \textit{GlobalTrans}, the rendered result in \myreffig{fig:scalar} corresponds well to the reference from a nearly opposite view.
\textit{Ours} also has smaller warping errors, which are visualized at the bottom of \myreffig{fig:scalar_real1}. Corresponding videos are presented in Sec. 1.2 of the supplemental webpage.

\subsection{Synthetic Scenes with Complex Lighting}\label{sec:SynthticTest}

In this section, we show results of synthetic fluid scenes rendered using Blender with a complex lighting combination of point lights, directional lights and an environment map. The density volume is rendered with strong and non-linear attenuation as dense smoke.
Since it is hard to approximate the lighting condition with point and ambient lights in this setting, the Global Transport method can hardly be applied to these scenes and is thus excluded in the comparison.
In these scenes, we have five cameras evenly distributed on a circle with the target fluid in the center.
All results are displayed with a dark green background, which is different to the background color used during training, and thus can be considered as being ``transparent'' with no density accumulated along the rays.

\begin{figure}
    \centering
    \includegraphics[width=0.95\linewidth]{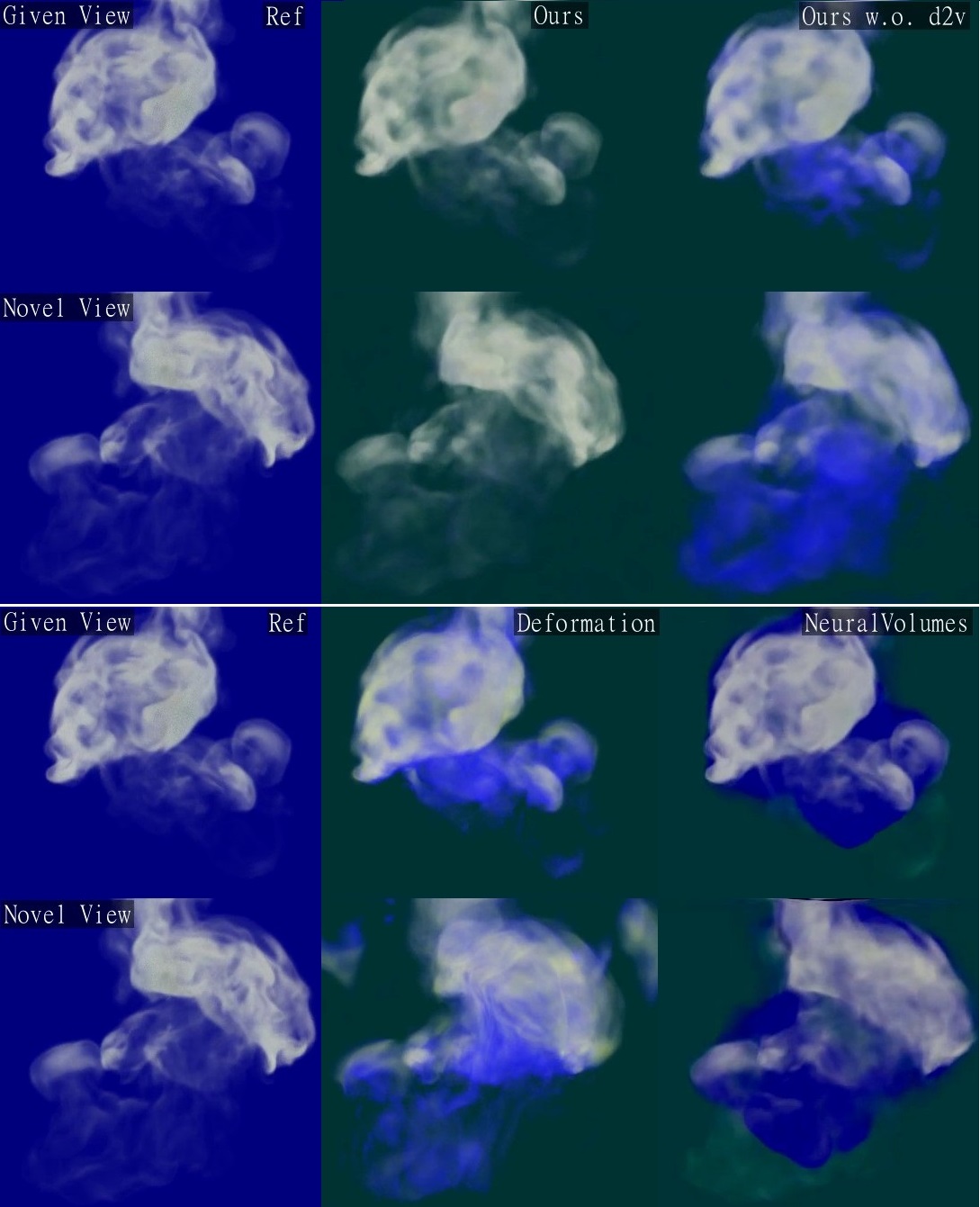}\\
    \includegraphics[width=0.95\linewidth]{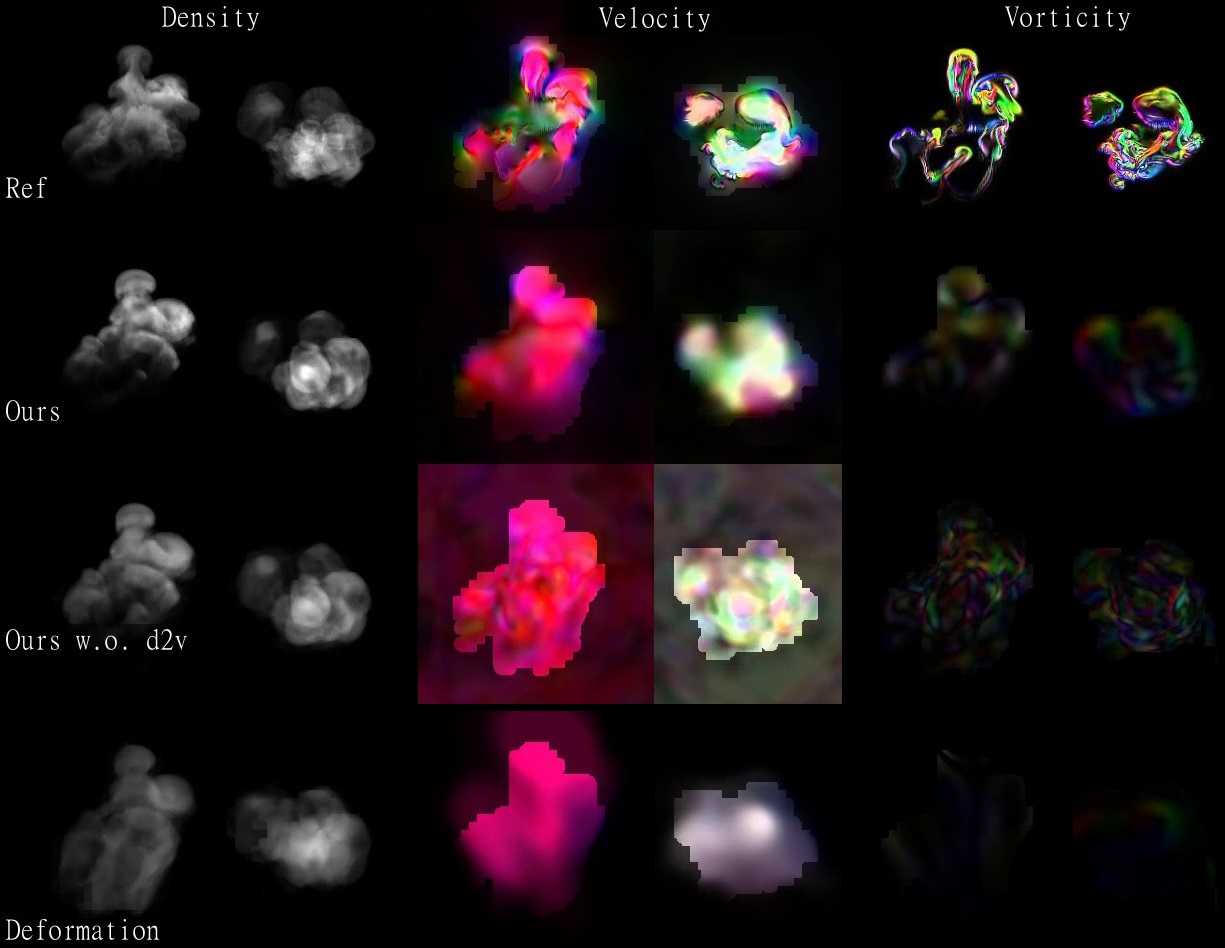}
    \caption{\footnotesize
    The plume scene. 
    \revise{Rendering results on the first two rows show that our method disentangles density and color successfully, while \textit{Ours w.o. d2v} has the color-bleeding artifact. \textit{Deformation} contains sharp edges in novel views due to its discontinuous deformation, while \textit{NeuralVolumes} suffers from ``ghost density''. Our velocity reconstruction in the third row is closest to the ground-truth.
    }    }
    \label{fig:plume}
\end{figure}

\begin{figure}[tp]\tiny
    \centering
    \begin{overpic}[width=0.95\linewidth]{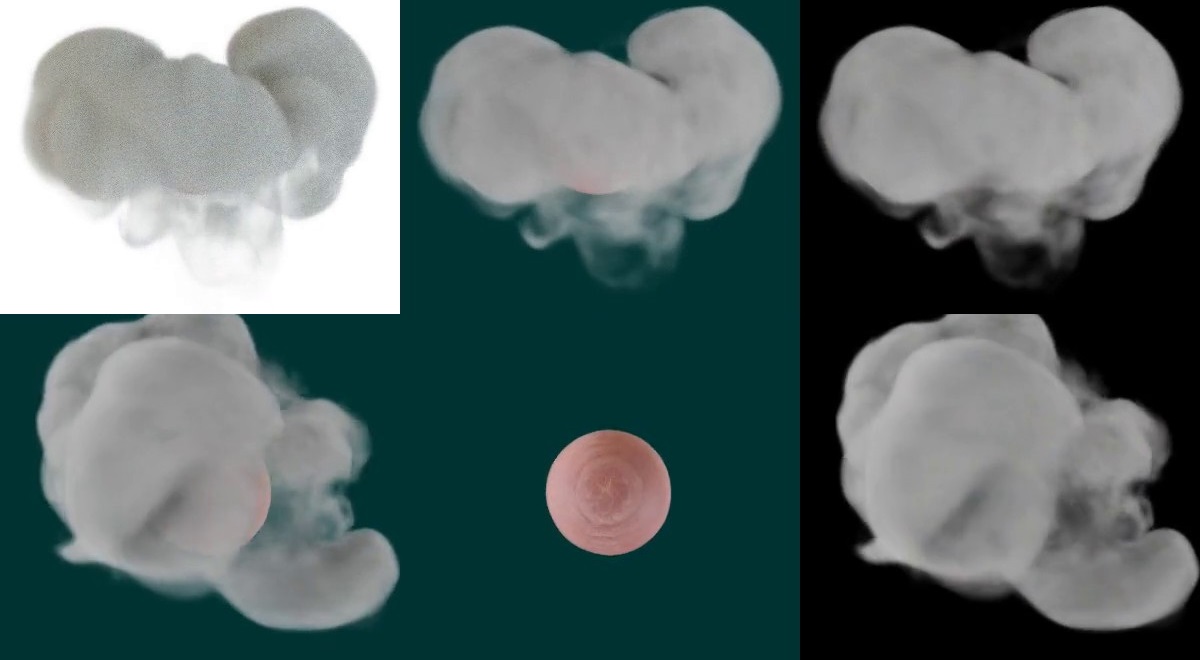}
    \put(2,30){Ref, A Novel View}
    \put(35,30){\color{white} Ours, Novel View 1}
    \put(68,30){\color{white} Our Fluid, Novel View 1}
    \put(2,3){\color{white}Ours, Novel View 2}
    \put(35,3){\color{white} Our Static, Novel View 2}
    \put(68,3){\color{white} Our Fluid, Novel View 2}
    \end{overpic}\\
    \begin{overpic}[width=0.95\linewidth]{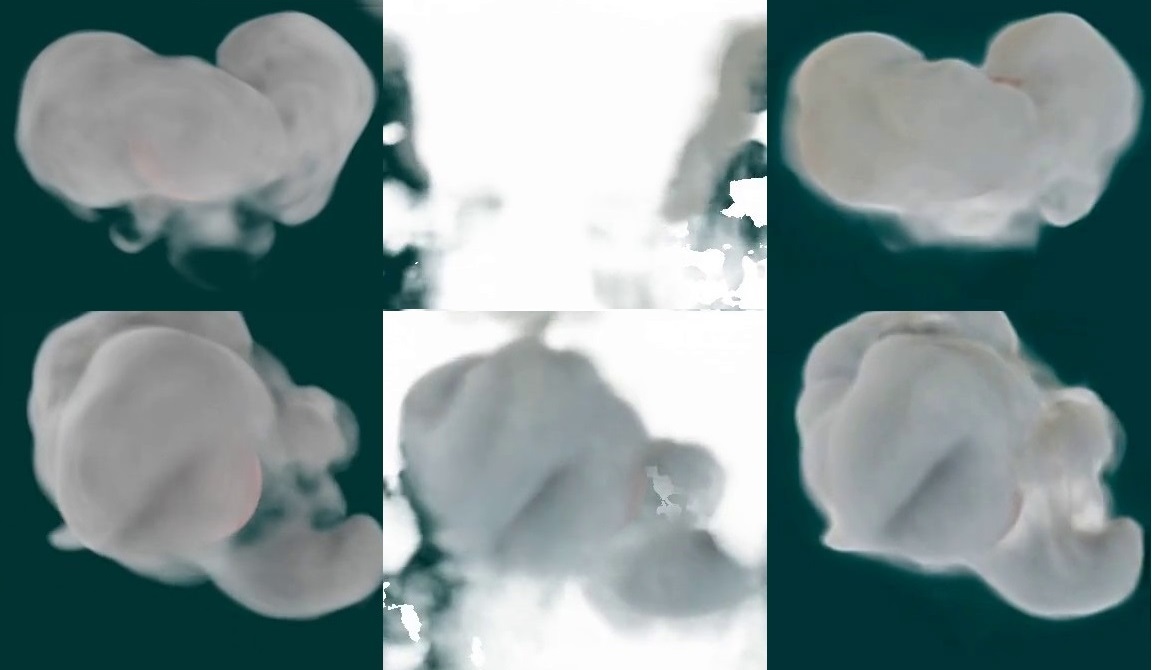}
    \put(2,33){\color{white} Ours w.o. d2v, Novel View 1}
    \put(35,33){ NeRF+T, Novel View 1}
    \put(68,33){\color{white} NeuralVolumes, Novel View 1}
    \put(2,2){\color{white} Ours w.o. d2v, Novel View 2}    \put(35,2){NeRF+T, Novel View 2}
    \put(68,2){\color{white} NeuralVolumes, Novel View 2}
    \end{overpic}\\
    \includegraphics[width=0.95\linewidth]{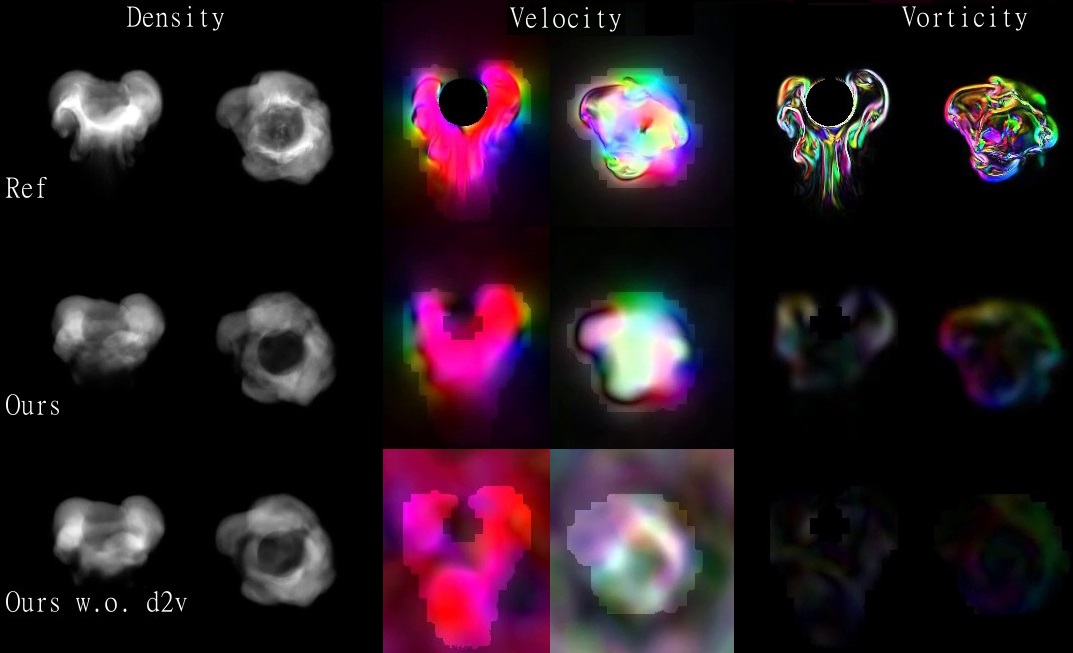}
    \caption{\footnotesize A plume scene with a sphere obstacle
    \revise{
    With a hybrid architecture, \textit{Ours} reconstruct static obstacles and dynamic fluids separately in an unsupervised manner, while \textit{NeuralVolumes} and \textit{NeRF+T} have ``ghost density''.
    Our reconstructed velocity closely follows the ground-truth, while \textit{Ours w.o. d2v} underestimates vorticity.
    }}
    \label{fig:sphere}
\end{figure}
\begin{figure*}[btp]\footnotesize
    \centering
    \begin{overpic}[width=0.595\linewidth]{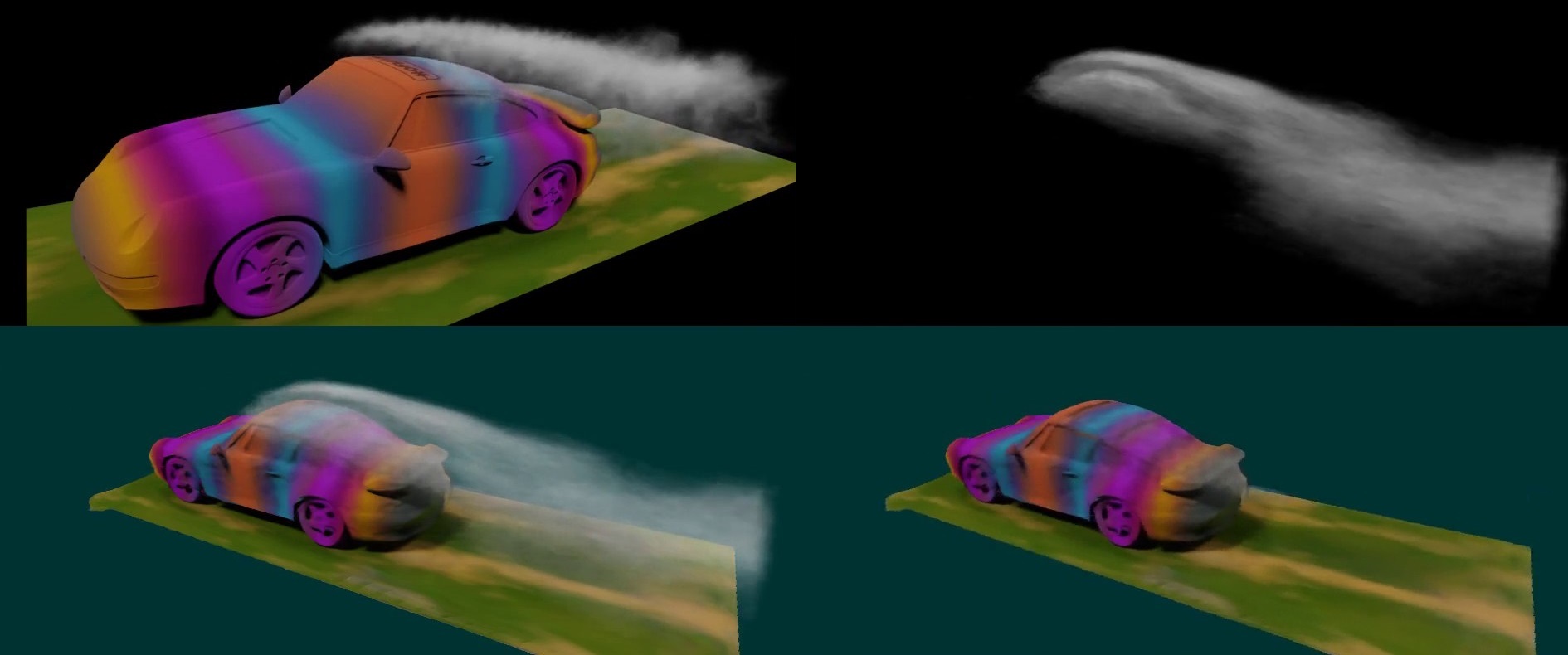}
    \put(1,39){\footnotesize \color{white} Ref, A Given View}
    \put(77,39){\footnotesize \color{white} Fluids, A Novel View}
    \put(75,18){\footnotesize \color{white} Static part, A Novel View}
    \put(1,18){\footnotesize \color{white}  Ours, A Novel View}
    \end{overpic}~
    \begin{overpic}[width=0.395\linewidth]{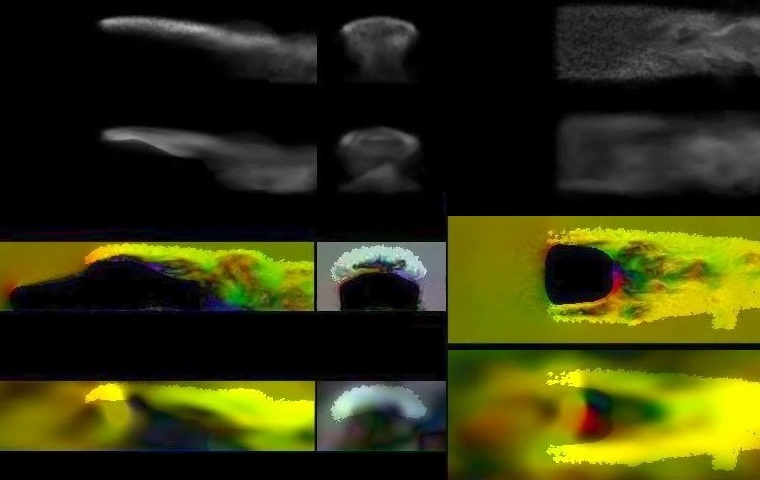}
    \put(1,1){\color{white}  Our velocity, side-front-top}
    \put(1,16){\color{white}  Ref velocity, side-front-top}
    \put(1,33){\color{white}  Our density, side-front-top}
    \put(1,50){\color{white}  Ref density, side-front-top}
    \end{overpic}
    \caption{Rendering and visualization of the radiance, density, and velocity fields reconstructed on the ``car'' scene. 
    \revise{Our method reconstructs the density distribution of the smoke and the complex geometry of the car using only 7 camera views.}}
    \label{fig:car}%
\end{figure*}

\paragraph{A Simple Plume Scene} 
We first test with a regular plume scene in resolution of $256^3$. 120 frames are used with a time step of 1.0.
While the ScalarFlow dataset has a black background, the blue color is used in this case, as shown by the given view in \myreffig{fig:plume}.
Due to the complex rendering setting with sparse views given, \textit{NeuralVolumes} shows strong ``ghost density'' in blue. This artifact is largely reduced in \textit{Ours w.o. d2v} with the help of $\mathcal{L}_{ghost}$, but is still visible at the bottom part with thin smoke. Our full model presents a good density estimation with ``ghost density'' hardly noticeable.
This improvement over \textit{Ours w.o. d2v} is mainly due to a more accurate velocity field.
Considering the thin smoke at the bottom part of \textit{Ours w.o. d2v}, a less accurate velocity field could warp them to a region expecting fluid density in white at some time, and warp them to a region expecting nothing or at least something in blue at some other time.
The temporal inconsistency results in a ``color bleeding'' artifact that is slightly visible in given views and more visible in novel views.

Some related works~\cite{tretschk2020nonrigid, Li2020NeuralSF} explore dynamic NeRF, but most of them focus on scenes with deformable surfaces and introduce constraints that are not appropriate for fluids. 
While we do not compare to these methods due to the largely different reconstruction target,
we train a \textit{Deformation} model to illustrate the impact of inappropriate constraints.
Learning a canonical spatial radiance field $F_{canon}:(x,y,z)\rightarrow (\textbf{c},\sigma)$ and a spatiotemporal deformation field $F_{deform}:(x,y,z,t)\rightarrow (x_{canon},y_{canon},z_{canon})$, \textit{Deformation} uses 
the same network architecture as ours.
Their pixels are rendered with 
$\Big(\mathbf{c}(h_k), \sigma(h_k)\Big) = F_{canon}\Big(F_{deform}(r_{ijt}(h_k))\Big)$.
As shown in \myreffig{fig:plume}, the \textit{Deformation} model achieves quality similar to \textit{Ours w.o. d2v} on the given view, but its density volume has unnatural stretches that are visible in novel views.
From the visualization of the density at the bottom left,
we can see the existence of ``ghost density'' at the bottom for the \textit{Deformation} model and \textit{Ours w.o. d2v}. 
While \textit{Deformation} does not produce velocity fields directly, we calculated one from $F_{deform}$ for visualization.
As shown on the bottom of \myreffig{fig:plume}, the velocity calculated from $F_{deform}$ is heavily constrained by the deformation and uniformly goes upward with almost zero vorticity. This relatively rigid deformation results in the stretches discussed above.
Our full model generates density and velocity volumes that can roughly match the reference.
The density in our result is more concentrated on the "surface" since the reference fluid has nonlinear attenuation, which is not given but can easily be extended if provided.
Videos of the plume scene are presented in Sec. 2.1 of the supplemental webpage.

\paragraph{Plume with a Regular Obstacle}
While previous scenes have fluids as the sole target, 
we test hybrid scenes with obstacles in the following.
The first hybrid scene has a regular obstacle in shape of a sphere.
The simulation resolution is $256^3$ and 148 frames are simulated with a time step of 1.0. The background is set as white during training.
As shown in the first row of \myreffig{fig:sphere}, based on our hybrid architecture and the deferred temporal component, our full model has the static and dynamic components successfully separated.
While our full model generates natural density volume without ``ghost density'', the result of \textit{Ours w.o. d2v} on the left of the second row is  slightly blurry.
\textit{NeuralVolumes} contains ``ghost density''. The baseline model of \textit{NeRF+T}, which is the original ReLU-based NeRF model with time as an extended dimension, has problems dealing with the white background and results in ``ghost density'' fulfilling the domain. Due to the occlusion of the ``ghost density'', there is a lack of supervision in the inner region and some density noise can be observed at novel views.
Visualizations of velocity and vorticity are shown at the bottom,
with ours closely resembling the reference.
Its vorticity is also much stronger than \textit{Ours w.o. d2v}.
The supplemental webpage presents the videos for this scene in Sec. 2.2.

\begin{figure}[tbp]
    \centering
    \includegraphics[width=0.95\linewidth]{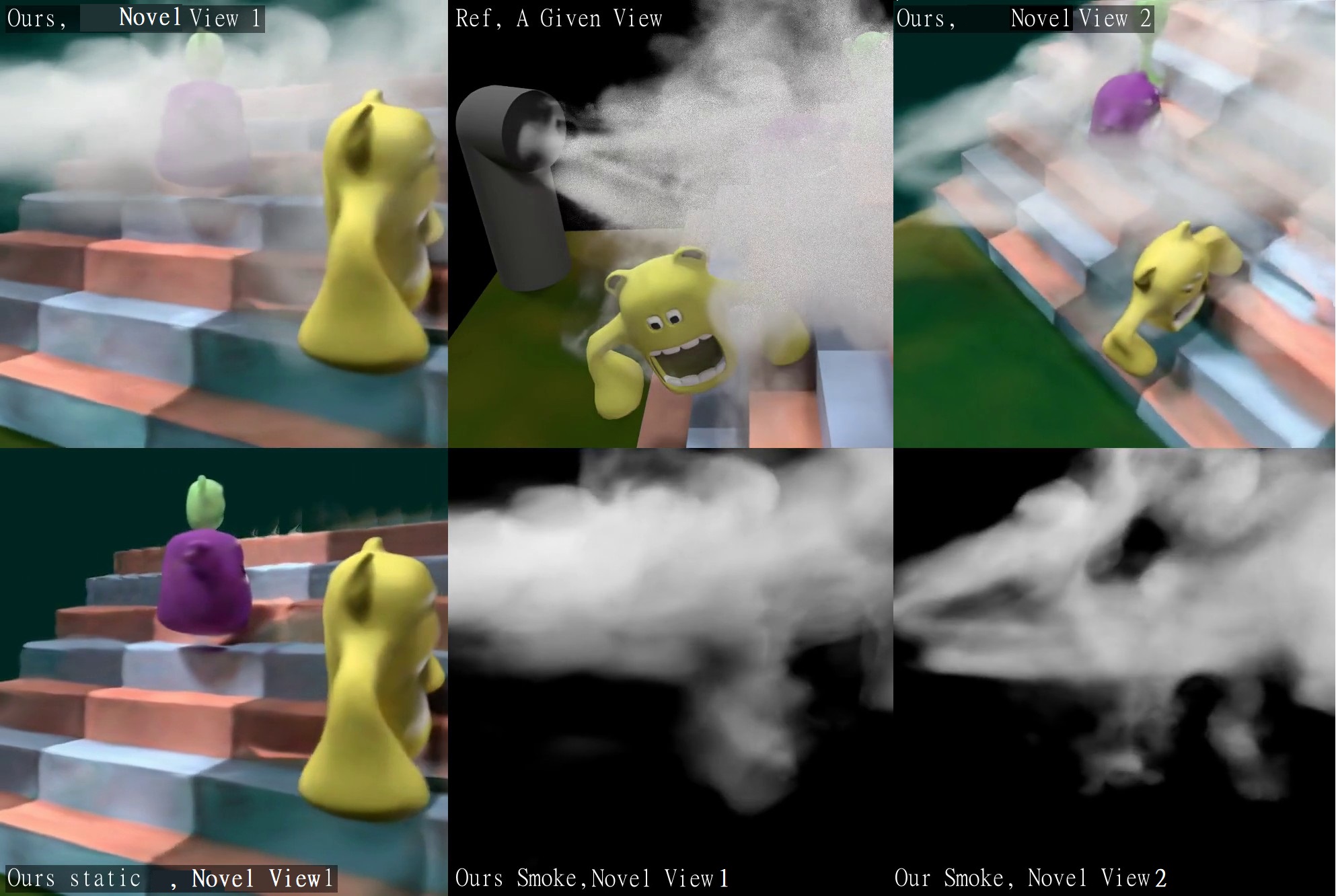}\\
    \includegraphics[width=0.95\linewidth]{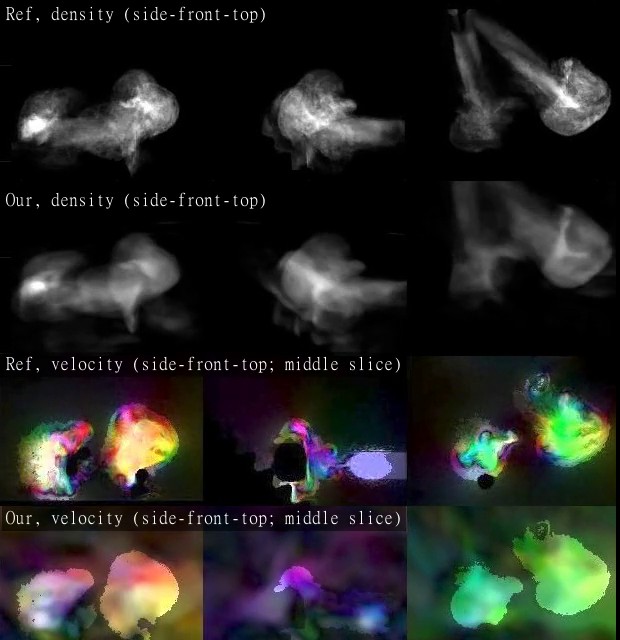}
    \caption{Rendering and visualization of the radiance, density, and velocity fields reconstructed on the ``game'' scene. Our velocity large resembles the complex reference.}
    \label{fig:game}
\end{figure}

\subsection{Complex Scenes with Arbitrary Obstacles}\label{sec:ObsTest}
At last, we test our algorithm in complex scenes with obstacles of arbitrary shape under lighting composited by point lights, directional lights, as well as an environment map.
The black color is used as the background during the training.
Besides the five evenly distributed cameras on a circle, we use two more cameras viewing from above since the geometry is very challenging in the following scenes. 

\paragraph{The Car Scene}

The \textit{Car} scene is simulated in resolution of $768\times192\times307$ for 148 steps with a time step of 1.0.
While other scenes only have one or two small inflow regions, the whole domain of this scene uses a strong free stream velocity, whose direction is visualized as yellow on the right of \myreffig{fig:car}.
We simulate a shallow sheet of smoke passing through the car from the front top.
The velocity field in this scene is particularly important for vehicle design, which can be used to calculate the drag force, the surface pressure, etc.
Using our algorithm, the geometry of the car and the density of the smoke are nicely and separately reconstructed.
While the reconstructed velocity roughly matches the reference, 
its vorticity is not as turbulent as the ground truth.
This is mainly because smoke density is mixed together due to the strong vorticity in all directions at the wake of the car and the observed density derivatives in space and time are not as high as it should be.
To improve on this particular case, it would be interesting to experiment with smoke in varying color or to apply frequency-based priors to constrain the vorticity.

\paragraph{The Game Scene}

In the \textit{Game} scene, dense smoke is coming out of a tube and hitting three monsters on the stairs. It is in resolution of $512\times432\times408$. 148 frames are simulated with a time step of 1.0. This scene represents a very difficult case due to the complex geometry, the occlusion caused by the close positioning of the monsters and stairs, the dense smoke hiding details inside, the strong motion at the interface of smoke and obstacles, and the moving shadow cast from the smoke.
Faced with all these difficulties, our method manages to separate the static and dynamic components very well. 
While the inner region of the reconstructed smoke density is a little blurry, the rendered images have reasonable details on the outer bound area of the smoke. 
The complex geometry of obstacles are reconstructed reasonably well in general.
The reconstructed velocity field can nicely resemble the complex ground truth, with smoke density accurately flowing around obstacles other than passing through.

\subsection{Results Summary and Limitations} 

To summarize, we have tested our algorithm on real and synthetic fluid scenes. For synthetic simulations, we have used buoyant and stream flow, with and without inflow, with and without obstacles in regular and arbitrary shapes. For the rendering, we have used dense and thin smoke, linear and nonlinear attenuation, simple and complex lighting conditions.

We observe consistent results provided by our model:
``Ghost density'' is successfully removed in density fields as an appropriate disentanglement of density and color.
The resulting velocity generally matches the reference and showing enhanced vorticity than purely PINN-based learning.
Additionally, our hybrid model separately reconstructs static and dynamic components without using additional manual labeling. 
The limitations of our method mainly pertain to non-linear aspects of the optimization process and slow training caused by the PDEs calculation via auto-differentiation.
While there is room for improvement on the high-frequency details of the reconstructed velocity field,
end-to-end estimation of velocity from images is a difficult task, and the proposed model-based supervision yields significant improvements.
\revise{Applying frequency-based supervision~\cite{yifan2022geometry} on the velocity would be an interesting future avenue.}
Also, the training of a single NeRF model is not efficient due to prohibitive queries of the neural networks. Training our algorithm with a hybrid architecture and a PINN-based velocity model on the dynamic data is around three times slower than that.
\revise{Training details including hyper-parameters and training time for each scene is provided in the supplemental material, for e.g., the Game scene takes 64 hours when using a single NVIDIA Quadro RTX 8000 GPU. }
With the recent progress in fast neural representation training ~\cite{yu2021plenoctrees, muller2022instant}, we anticipate this limitation to be resolved in the near future.
\revise{%
Besides code optimization and the use of multiple GPUs, breaking down large-scale scenes in space can make their reconstructions more efficient.
Our fluid reconstruction can be used to enhance physical understanding from videos. 
The reconstructed fields can also be used in graphics applications. 
In contrast with forward simulation methods that allow fluid animations to be designed using initial conditions and physical parameters, reconstruction methods like ours allow users to generate fluid phenomena from video captures. In order to present fluid animations with fine-level detail, it would be interesting to apply our approach together with detail synthesis~\cite{you2018tempoGAN} or fluid guiding~\cite{forootaninia2020frequency}.
}

\section{Conclusions}

We have introduced an optimization-based algorithm that is able to reconstruct continuous fluid fields end-to-end from a sparse set of video frames based on the developed spatio-temporal neural representation for dynamic fluid flow and the underlying physics-informed learning mechanisms.
To the best of our knowledge, this is the first method to 
allow flow motion to be reconstructed from image captures of hybrid scenes with both fluid and arbitrary obstacles, while being agnostic to initial, boundary, or lighting conditions.
In our optimization framework, we jointly apply supervision from images, physical priors, as well as data prior encoded as a pre-trained model.
With the comprehensive supervisions, our method exhibits stable effectiveness and strong flexibility on a wide range of scenes.

We see our method as an crucial step towards capturing and analyzing real fluid phenomena with relaxed constraints, e.g. allowing captures under changing illumination. With the advantage of handling scenes with unknown obstacles and lighting, we are especially interested in exploring in-the-wild fluid capture as well as more elaborate fluid-obstacle interactions in the future.

\begin{acks}
The project was supported by the ERC Consolidator Grant 4DRepLy (770784). Mengyu Chu and Lingjie Liu were supported by Lise Meitner Postdoctoral Fellowships.
\end{acks}
{\small
	\bibliographystyle{ACM-Reference-Format}
	\bibliography{egbib}
}
\end{document}